\documentclass[%
twocolumn,
 amsmath,amssymb,
 aps,
 prl,
]{revtex4}
\usepackage{dcolumn}% Align table columns on decimal point
\usepackage{bm}% bold math
\usepackage{graphicx}
\usepackage{comment}
\begin{document}

%\preprint{}

\title{Time between the maximum and the minimum of a stochastic process}% Force line breaks with \\

\author{Francesco Mori}
\address{LPTMS, CNRS, Univ.  Paris-Sud,  Universit\'e Paris-Saclay,  91405 Orsay,  France}

\author{Satya N. Majumdar}
\address{LPTMS, CNRS, Univ.  Paris-Sud,  Universit\'e Paris-Saclay,  91405 Orsay,  France}

\author{Gr\'egory Schehr}
\address{LPTMS, CNRS, Univ.  Paris-Sud,  Universit\'e Paris-Saclay,  91405 Orsay,  France}

\date{\today}

%\begin{abstract}
%The celebrated \textit{arcsine law} describes the probability density function (p.d.f.) of the time of the maximum of Brownian motion. However, it is known that the time of maximum $t_{max}$ and the time of minimum $t_{min}$ are not independent. The \textit{arcsine law} does not allow us to describe quantities that involve both $t_{max}$ and $t_{min}$. For Brownian motion and Brownian bridge evolving for a time $T$, we present exact formulae for the p.d.f. of the time difference $u=t_{min}-t_{max}$. This can be written in the scaling form $P(u|T)=1/Tf(u/T)$. Using this result, we compute the covariance of $t_{max}$ and $t_{min}$, showing that they are anti-correlated. Our results are confirmed by extensive numerical simulations.
%\end{abstract}
\begin{abstract}
We present an exact solution for the probability density function $P(\tau=t_{\min}-t_{\max}|T)$ of the time-difference between the minimum and the maximum of a one-dimensional Brownian motion of duration $T$. We then generalise our results to a Brownian bridge, i.e. a periodic Brownian motion of period $T$. We demonstrate that these results can be directly applied to study the position-difference between the minimal and the maximal height of a fluctuating $(1+1)$-dimensional Kardar-Parisi-Zhang interface on a substrate of size $L$, in its stationary state. We show that the Brownian motion result is universal and, asymptotically, holds for any discrete-time random walk with a finite jump variance. We also compute this distribution numerically for L\'evy flights and find that it differs from the Brownian motion result.    
\end{abstract}

\maketitle

%\tableofcontents
\newpage
\begin{figure}[t]
\includegraphics[width=\linewidth]{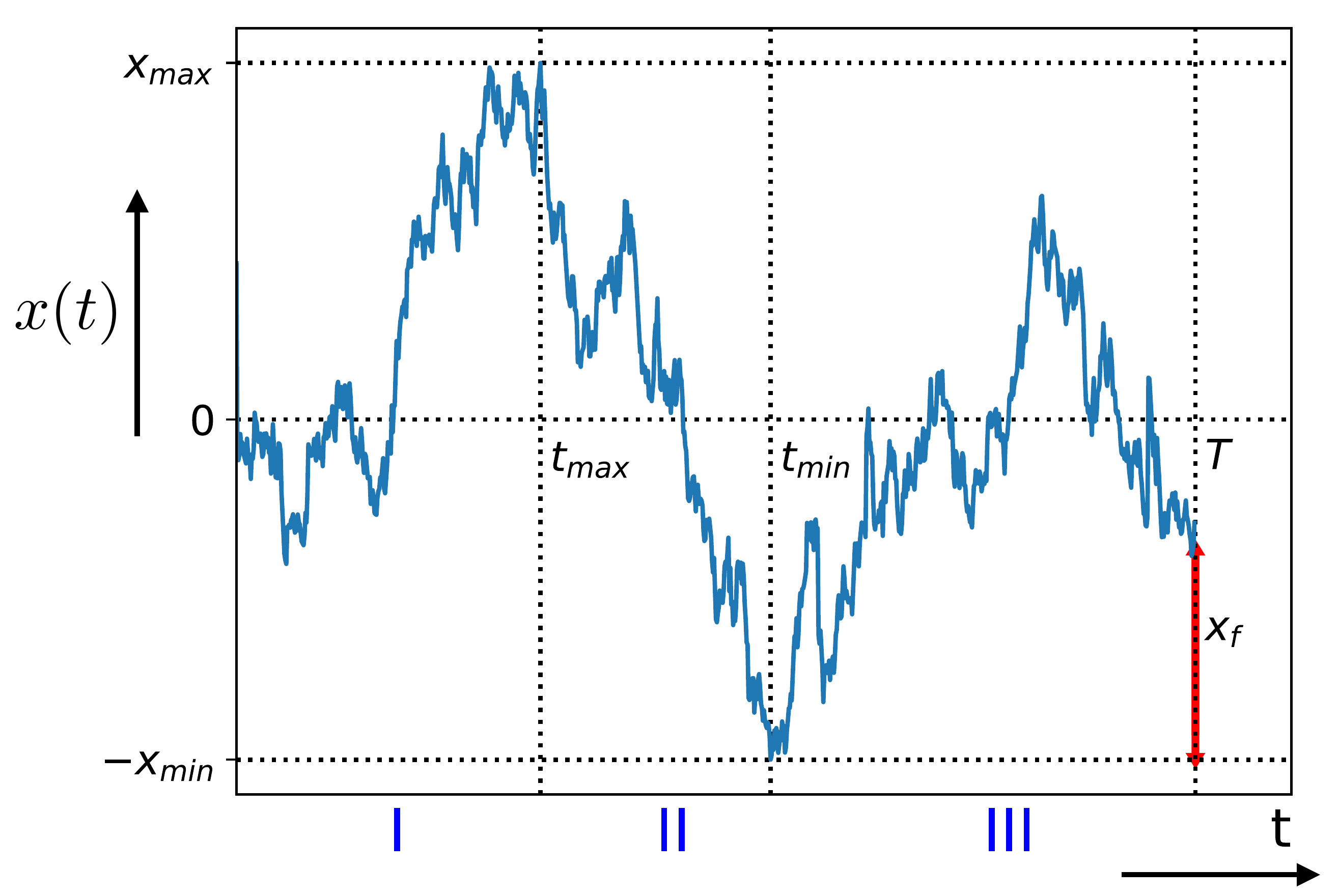} % Here is how to import EPS art
\caption{\label{fig:brownian} A typical trajectory of a Brownian motion $x(t)$ during the time interval $[0,T]$, starting from $x(0)=0$. The global maximum $x_{\max}$ occurs at time $t_{\max}$ and the global minimum $-x_{\min}$ at $t_{\min}$. The final position $x(T)$, measured with respect to $-x_{\min} \leq 0$ is denoted by $x_{\rm f}$. The total time interval $[0,T]$ is divided into three segments: $[0,t_{\max}]$ (I), $[t_{\max}, t_{\min}]$ (II) and $[t_{\min}, T]$ (III), for the case $t_{\min}>t_{\max}$.}
\end{figure}

The properties of extremes of a stochastic process/time series of a given duration $T$ are of fundamental importance in describing a plethora of natural phenomena~\cite{Gumbel,katz02,katz05,Krug07,NK11}. For example, this time series may represent the amplitude of earthquakes in a specific seismic region, the amount of yearly rainfall in a given area, the temperature records in a given weather station, etc. The study of extremes in such natural time series have gained particular relevance in the recent context of global warming in climate science~\cite{RP06,RC2011,Christ13,WHK14,MBK19}. Extremal properties also play an important role for stochastic processes that are outside the realm of natural phenomena. For example, in finance, a natural time series is the price of a stock for a given period~\cite{BP2000,yor01,embrechts13,Challet17}. Knowing the maximum or the minimum value of a stock during a fixed period 
is obviously important, but an equally important question is when does the maximum (minimum) value of the stock occurs within this period $[0,T]$. Let $t_{\max}$ and $t_{\min}$ denote these times (see e.g. Fig. \ref{fig:brownian}). For a generic stochastic process, computing the statistics of $t_{\max}$ and $t_{\min}$ is a fundamental and challenging problem. The simplest and the most ubiquitous stochastic process is the one-dimensional Brownian motion (BM) of a given duration $T$, for which the probability distribution function (PDF) of $t_{\max}$ can be computed exactly~\cite{levy40,feller50,SA53,morters10} 
\begin{equation}\label{levy}
P(t_{\max}|T)=\frac{1}{\pi\sqrt{t_{\max}(T-t_{\max})}} \;, \; 0\leq t_{\max} \leq T \;.
\end{equation}
The cumulative distribution ${\rm Proba.}\,(t_{\max}<t|T) = \frac{2}{\pi} \sin^{-1}(\sqrt{t/T})$ is known as the celebrated {\it arcsine law} of L\'evy \cite{levy40}. By the symmetry of the BM, the distribution of $t_{\min}$ is also described by the same arcsine law (\ref{levy}). In the past decades, the statistics of $t_{\max}$ has been studied for a variety of stochastic processes, going beyond BM. Examples include BM with drift \cite{majumdar08,drift}, constrained BM \cite{randon-furling07,randon-furling08,schehr10} (such as the Brownian bridge and the Brownian excursion, see also \cite{schehr10, PLDM03} for real space renormalization group method), Bessel processes \cite{schehr10}, L\'evy flights \cite{SA53,majumdar10}, random acceleration process \cite{rosso10}, fractional BM \cite{delorme16,tridib18}, run-and-tumble particles \cite{anupam_rtp}, etc. The statistics of $t_{\max}$ has also been studied in multi-particle systems, such as $N$ independent BM's \cite{comtet10} as well as for $N$ vicious walkers \cite{rambeau11}. 
Moreover, the arcsine law and its generalisations have been applied in a variety of situations such as in disordered systems \cite{MRZ10}, stochastic thermodynamics \cite{barato18}, finance \cite{dale80,baz04} and sports \cite{clauset15}. 

While the marginal distributions of $t_{\max}$ and $t_{\min}$ are given by the same arcsine law in Eq. (\ref{levy}) for a BM due its symmetry, 
one expects that $t_{\max}$ and $t_{\min}$ are strongly correlated. For example, if the maximum occurs at a certain time, it is unlikely that the minimum occurs immediately after or before. These anti-correlations are encoded in the joint distribution $P(t_{\min},t_{\max}|T)$ which does not factorise into two separate arcsine laws. These anti-correlations also play an important role for determining the statistics of another naturally important observable, namely the time-difference between $t_{\min}$ and $t_{\max}$: $\tau = t_{\min} - t_{\max}$. In the context of finance where the price of a stock is modelled by the exponential of a BM, $\tau$ represents the time-difference between the occurrences of the minimum and the maximum of the stock price. For instance, if $t_{\max} < t_{\min}$ as in Fig. \ref{fig:brownian}, an agent would typically sell her/his stock when the price is the highest and then wait an interval $\tau$ before re-buying the stock, because the stock is the cheapest at time $t_{\min}$. Hence a natural question is: how long should one wait between the buying and the selling of the stock? In other words, one would like to know the PDF $P(\tau|T)$ 
of the time difference $\tau$. To calculate this PDF, we need to know the joint distribution $P(t_{\min},t_{\max}|T)$. Computing this joint distribution and the PDF $P(\tau|T)$ for a stochastic process is thus a fundamentally important problem. 

\begin{figure}[t] 
   \includegraphics[width=\linewidth]{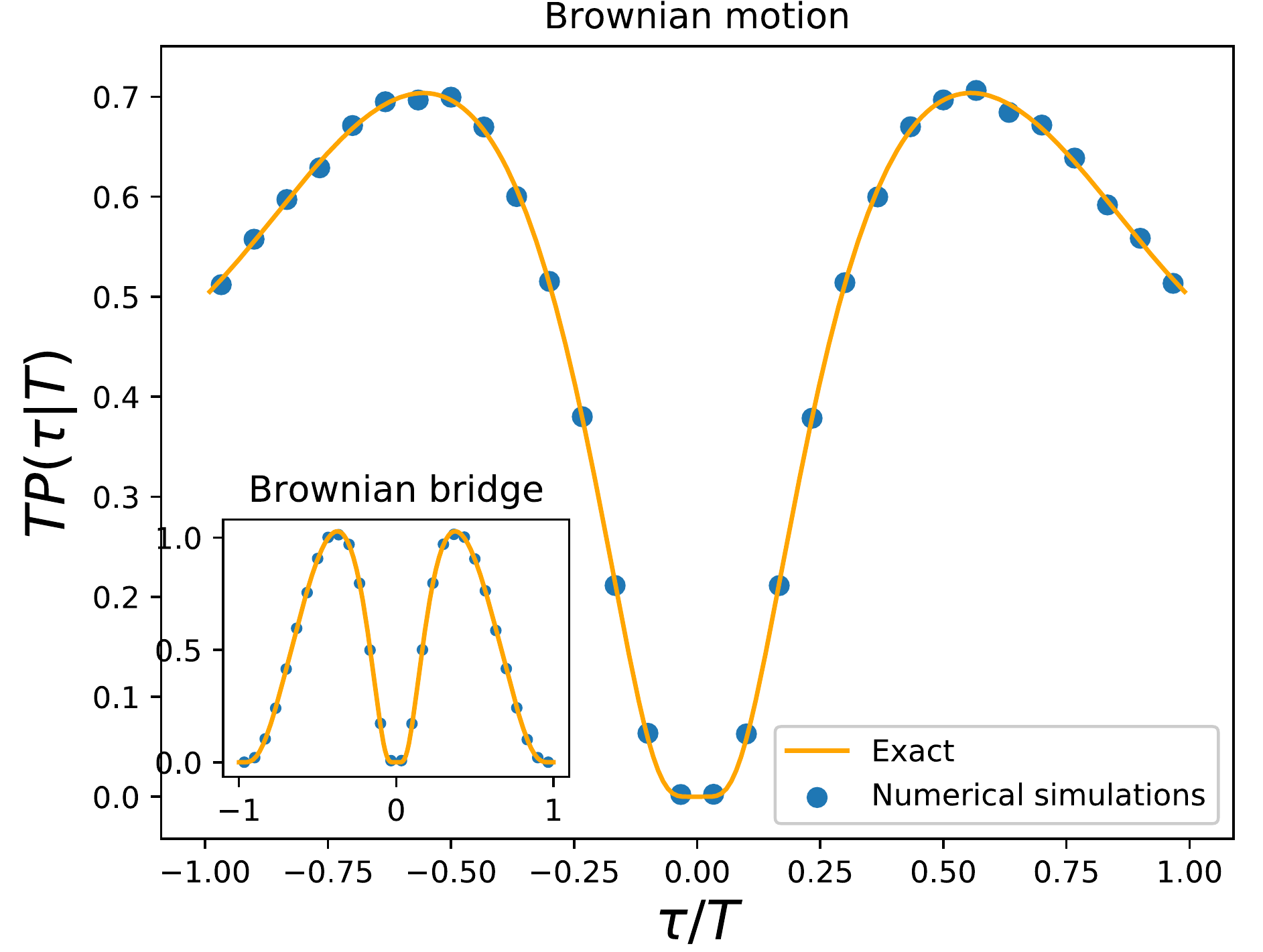}
    \caption{The scaled distribution $T\, P(\tau|T)$ plotted as a function of $\tau/T$ for the BM (the solid line corresponds to the exact scaling function $f_{\rm BM}(y)$ in Eq. (\ref{eq:f_bm}), while the filled dots are the results of simulations). {\bf Inset}: the same scaled distribution for the  Brownian bridge where the exact scaling function $f_{\rm BB}(y)$ is given in Eq. (\ref{eq:f_bb}).}
    \label{fig:numeric}
\end{figure}

\begin{figure}[t]    
    \includegraphics[width=\linewidth]{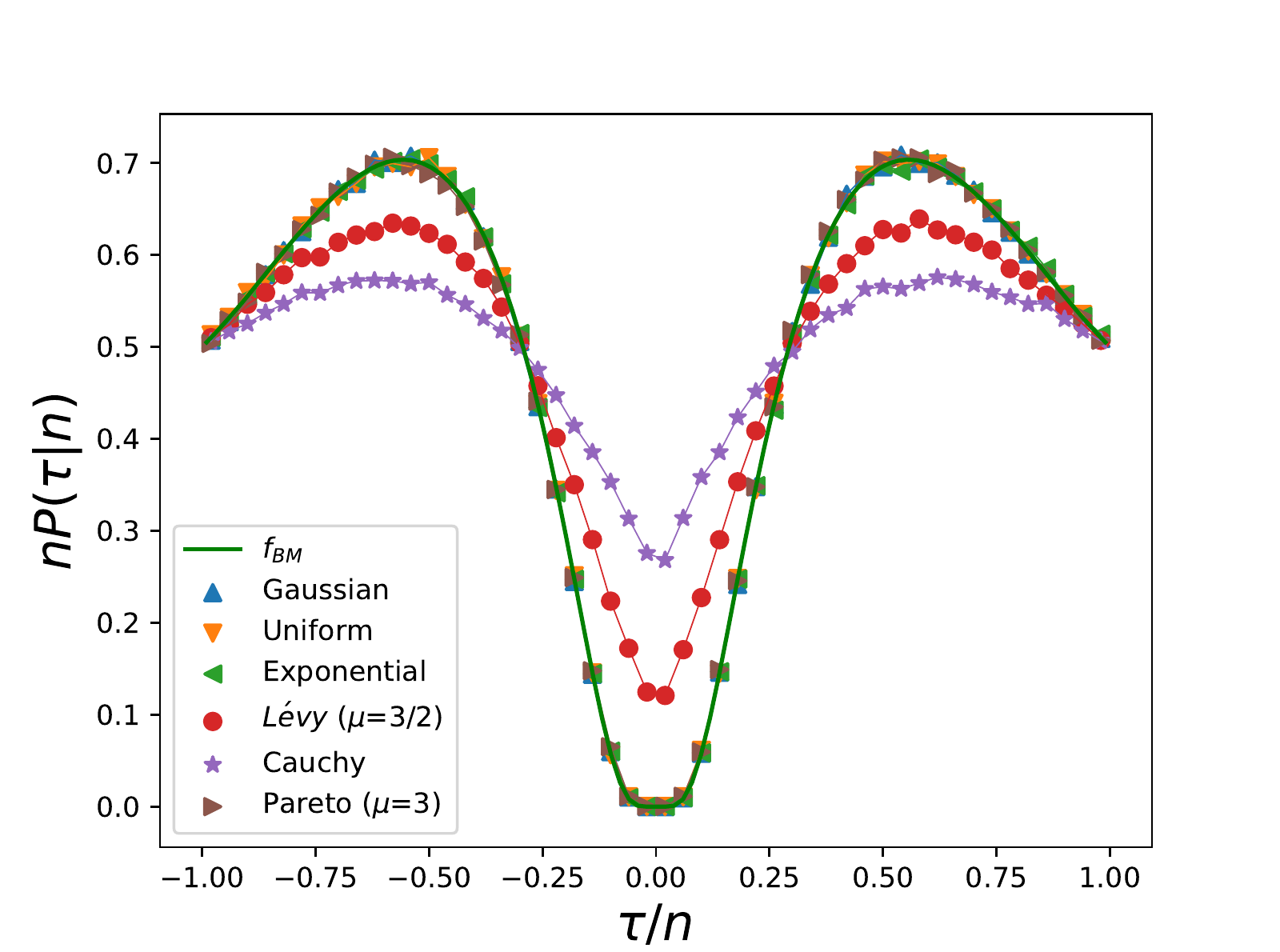} 
    \caption{The scaled distribution $n P(\tau|n)$ versus $\tau/n$ for RWs with different jump distributions. The Gaussian, uniform, double-exponential and Pareto (with index $\mu=3$) jump distributions, all of which have a finite variance, collapse onto the Brownian scaling function $f_{\rm BM}(y)$ shown by the solid (green) line. For L\'evy flights with index $\mu = 3/2$ and $\mu = 1$ (Cauchy distribution), the scaling function $f_{\mu}(y)$ depends on $\mu$ (except at the endpoints $y = \pm 1$ where $f_{\mu}(\pm 1) = 1/2$ seems to be universal for all $0<\mu\leq 2$).}
    \label{fig:numeric_rw}
\end{figure}

In this Letter, by using a path-integral method, we compute exactly the joint distribution $P(t_{\min},t_{\max}|T)$ for a BM of a given duration $T$. This joint distribution does not depend on the initial position $x_0$. Hence, without loss of generality, we set $x_0 = 0$. We also 
generalise our results to the case of a Brownian bridge (BB), which is a periodic BM of period $T$. Using these results, we first compute exactly the covariance function $\operatorname{cov}(t_{\min},t_{\max}) = \langle t_{\min} t_{\max} \rangle - \langle t_{\min} \rangle \langle t_{\max}\rangle$ that quantifies the anti-correlation between $t_{\max}$ and $t_{\min}$. We find  $\operatorname{cov}_{\rm BM}(t_{\min},t_{\max})=-\frac{7\zeta(3)-6}{32}T^2\approx -0.0754\textit{   } T^2$ for a BM ($\zeta(z)$ being the Riemann zeta function) while $\operatorname{cov}_{\rm BB}(t_{\min},t_{\max})=-\frac{\pi^2-9}{36}T^2\approx -0.0241\textit{   } T^2$ for a BB. 
%\begin{eqnarray}
%\operatorname{cov}_{\rm BM}(t_{\min},t_{\max})&=&-\frac{7\zeta(3)-6}{32}T^2\approx -0.0754\textit{   } T^2 \label{cov_BM} \;\; \\
%\operatorname{cov}_{\rm BB}(t_{\min},t_{\max})&=&-\frac{\pi^2-9}{36}T^2\approx -0.0241\textit{   } T^2 \label{cov_BB} \;,
%\end{eqnarray}
%where $\zeta(z)$ is the Riemann zeta function. 
From this joint distribution, we also compute exactly the PDF $P(\tau|T)$ for the BM as well as for the BB. We show that for a BM of duration $T$, starting at some fixed position, $P(\tau|T)$ has a scaling form for all $\tau$ and $T$: $P(\tau|T)=(1/T)\,f_{\rm BM}(\tau/T)$ where the scaling function $f_{\rm BM}(y)$ with $-1\leq y \leq 1$ is given by
\begin{eqnarray}\label{eq:f_bm}
f_{\rm BM}(y)=\frac{1}{|y| }\sum_{m=1}^{\infty}(-1)^{m+1}\tanh^2\left(\frac{m\pi}{2}\sqrt{\frac{|y|}{1-|y|}}\right).
\end{eqnarray}
Clearly $f_{\rm BM}(y)$ is symmetric around $y=0$, but is non-monotonic as a function of $y$ (see Fig.~\ref{fig:numeric} where we also compare it with numerical simulations, finding excellent agreement). This function has asymptotic behaviors 
\begin{eqnarray}\label{asymptote_bm}
f_{\rm BM}(y) \underset{y \to 0}{\approx} \frac{8}{y^2} e^{-\frac{\pi}{\sqrt{y}}} \;\;\; , \;\;\; f_{\rm BM}(y) \underset{y\to 1}{\approx} \frac{1}{2}  \;.
\end{eqnarray}
For a BB, we get
\begin{equation}\label{eq:f_bb}
f_{\rm BB}(y)=3\,(1-|y|)\sum_{m,n=1}^{\infty}\frac{(-1)^{m+n}m^2n^2}{\left[m^2|y|+n^2(1-|y|)\right]^{5/2}} \;,
\end{equation}
which is again symmetric around $y=0$ (see the inset of Fig. \ref{fig:numeric}) and has the asymptotic behaviors
\begin{eqnarray}\label{asymptote_bb}
f_{\rm BB}(y)\underset{y \to 0}{\approx}\frac{\sqrt{2}\pi^2}{y^{\frac{9}{4}}}e^{-\frac{\pi}{\sqrt{y}}} \;, \; f_{\rm BB}(y) \underset{y\to1}\approx\frac{\sqrt{2}\,\pi^2}{(1-y)^{\frac{5}{4}}}e^{-\frac{\pi}{\sqrt{1-y}}} .
\end{eqnarray}
Next, we demonstrate that these scaling functions are universal in the sense of Central Limit Theorem. Indeed, consider a time series of $n$ steps generated by the positions of a random walker evolving via the Markov rule
\begin{eqnarray}
x_{k} = x_{k-1}+\eta_k \;, \label{def_RW}
\end{eqnarray}
starting from $x_0=0$, where $\eta_k$'s are independent and identically distributed random jump variables, each drawn from a symmetric and continuous PDF $p(\eta)$. For all such jump distributions with a finite variance $\sigma^2 = \int \eta^2 p(\eta)\, d\eta$, we expect that, for large $n$, the corresponding PDF of the time difference $\tau = t_{\min} - t_{\max}$ would converge to the same scaling function $f_{\rm BM}(y)$ as the BM. Similar result holds for the BB as well. We confirm this universality by an exact computation for $p(\eta)= (1/2)e^{-|\eta|}$ and numerically for other distributions with finite $\sigma$ (see Fig. \ref{fig:numeric_rw}). 
In contrast, for L\'evy flights $\sigma^2$ is divergent since $p(\eta)$ has a heavy tail $p(\eta) \sim 1/|\eta|^{\mu+1}$ for large $\eta$, 
with index $0<\mu<2$. In this case, the PDF of $\tau$ is characterised  by a different scaling function parameterised by $\mu$ that we compute numerically. Finally, we show how our results can be applied to study similar observables for heights in the stationary state of a Kardar-Parisi-Zhang (KPZ) or Edwards-Wilkinson (EW) interfaces in one-dimension on a finite substrate of size~$L$.

\par
\emph{Brownian motion.} We start with a BM $x(t)$ over the time interval $t \in [0,T]$, starting at $x(0) = 0$ as in Fig.~\ref{fig:brownian}. Let $x_{\max} = x(t_{\max})$ and $-x_{\min}=x(t_{\min})$ denote the magnitude of the maximum and the minimum in $[0,T]$. Our strategy is to first compute the joint distribution $P(x_{\min},x_{\max},t_{\min},t_{\max}|T)$ of these four random variables and then integrate out $x_{\max}$ and $x_{\min}$ to obtain the joint PDF $P(t_{\min}, t_{\max}|T)$. We note that while the actual values $x_{\max}$ and $x_{\min}$ do depend on the starting position $x_0$, their locations $t_{\max}$ and $t_{\min}$ are independent of $x_0$, since it corresponds to a global shift in the position but not in time. The joint distribution of $x_{\max}$ and $x_{\min}$ can be computed by integrating out $t_{\max}$ and $t_{\min}$ (see e.g. \cite{KMS13,MSS16}). Without any loss of generality, we will assume that $t_{\max}<t_{\min}$ (the case $t_{\min} < t_{\max}$ can be analysed in the same way). The grand joint PDF $P(x_{\min},x_{\max},t_{\min},t_{\max}|T)$
can be computed by decomposing the time interval $[0,T]$ into three segments: (I) $[0,t_{\max}]$, (II) $[t_{\max}, t_{\min}]$ and (III) $[t_{\min},T]$. In the first segment (I), the trajectory starts at $0$ at time $t=0$, arrives at $x_{\max}$ at time $t_{\max}$ and stays inside the space interval $x(t) \in [-x_{\min}, x_{\max}]$ during $[0,t_{\max}]$ (see Fig. \ref{fig:brownian}). In the second segment (II), the trajectory starts at $x_{\max}$ at time $t_{\max}$ and arrives at $-x_{\min}$ at time $t_{\min}$ and stays inside the box $[-x_{\min}, x_{\max}]$. Finally, in the third segment (III), the trajectory starts at $-x_{\min}$ at time $t_{\min}$ and stays inside the box $[-x_{\min}, x_{\max}]$ during $[t_{\min},T]$. Clearly the trajectory stays inside the box $[-x_{\min}, x_{\max}]$ because, by definition, it can neither exceed its maximum value $x_{\max}$, nor can it go below its minimum $-x_{\min}$. Let us also denote by $M = x_{\min} + x_{\max} \geq 0$. 

To enforce the trajectory to stay inside the box, one needs to impose absorbing boundary condition at both $x_{\max}$ and $-x_{\min}$. However, for a continuous time Brownian motion, it is well known that one can not simultaneously impose absorbing boundary condition and also require the trajectory to arrive exactly at the absorbing boundary at a certain time. One way to get around this problem is to introduce a cut-off $\epsilon$ such that the actual values of the maximum and the minimum are respectively $x_{\max} - \epsilon$ and $-x_{\min}+\epsilon$, see Fig. 1 in the Supp. Mat. (SM) \cite{supmat}. One then computes the probability of the trajectory for a fixed $\epsilon$ and eventually takes the limit $\epsilon \to 0$. We use the Markov property of the process to express the total probability of the trajectory as the product of the probabilities of the three individual time segments. These probabilities can be expressed in terms of a basic building block or Green's function defined as follows. Let $G_{M}(x,t|x_0,t_0)$ denote the probability density for a BM, starting at $x_0$ at $t_0$, to arrive at $x$ at time $t$ while staying inside the box $x(t) \in [0,M]$ during $[t_0,t]$. Note that $x,x_0 \in [0,M]$. This Green's function can be computed by solving the Fokker-Planck equation, $\partial_t G_M = (1/2) \partial_x^2 G_M$, with absorbing boundary conditions $G_M = 0$ at both $x=0$ and $x=M$~\cite{Redner,Risken}
 \begin{equation}\label{eq:g} 
G_M=\frac{2}{M}\sum_{n=1}^{\infty}\sin\left(\frac{n\pi x}{M}\right)\sin\left(\frac{n \pi x_0}{M}\right)e^{-\frac{n^2\pi^2}{2M^2}(t-t_0)}.
\end{equation}
We first shift the origin in Fig. \ref{fig:brownian} to $-x_{\min}$. The probability of the trajectory in the three segments is then proportional respectively to (I) $G_{M}(M-\epsilon,t_{\max}|x_{\min},0)$, (II) $G_{M}(\epsilon,t_{\min}|M-\epsilon,t_{\max})$ and (III) $\int_{0}^{M} \hspace*{-0.cm}G_{M}(x_{\rm f},T|\epsilon,t_{\min}) \,dx_{\rm f}$, where, in the last segment, we integrate over the final position  
$x_{\rm f} = x(T) + x_{\min}$ of the trajectory inside the box $[0,M]$ (see Fig. \ref{fig:brownian}). Hence $P(x_{\min},x_{\max},t_{\min},t_{\max}|T)$ is given, up to an overall normalization factor, by the product of three individual pieces. Using the Green's function in Eq. (\ref{eq:g}), one can compute explicitly this grand PDF and finally integrate over $x_{\max}$ and $x_{\min}$ to obtain the joint PDF $P(t_{\min}, t_{\max}|T)$ [see SM~\cite{supmat} for details]. From the latter, we compute the covariance of $t_{\min}$ and $t_{\max}$ explicitly, as given above. 
Also, by integrating the joint PDF $P(t_{\min}, t_{\max}|T)$ over $t_{\max}$ and $t_{\min}$ with $t_{\min} - t_{\max} = \tau$ fixed, we compute explicitly the PDF $P(\tau|T)$ and find that it has the scaling form given in Eq. (\ref{eq:f_bm})~\cite{supmat}.   

\emph{Discrete time random walks and L\'evy flights.} It is natural to ask whether the PDF $P(\tau | T)$ of the continuous-time BM derived above holds for discrete-time random walks/L\'evy flights defined in Eq. (\ref{def_RW}). For such a discrete-time walk, let $n_{\max}$ ($n_{\min}$) denote the time at which the maximum (minimum) is reached. Remarkably, Sparre Andersen showed that 
the marginal distribution $P(n_{\max}|n)$ [equivalently $P(n_{\min}|n)$] is completely universal for all $n$, i.e. is the same for any symmetric jump PDF $p(\eta)$ \cite{SA53} -- thus it is identical both for random walks with finite jump variance and L\'evy flights. In particular, for large $n$, $P(n_{\max}|n)$ converges to the arcsine law in Eq. (\ref{levy}), with $T$ replaced by $n$ and $t_{\max}$ replaced by $n_{\max}$. Does this universality hold  also for the PDF of the time difference $\tau = n_{\min} - n_{\max}$? To investigate this question, we computed $P(\tau|n )$ exactly for the special of a double exponential jump PDF $p(\eta) = (1/2) e^{-|\eta|}$. The details are left in the SM \cite{supmat} and here we just outline the main results. We find that, for large $n$, $P(\tau |n)$ converges to the scaling form $P(\tau|n) \approx (1/n) f_{\rm exp}(\tau/n)$ where $f_{\rm exp}(y)$ satisfies the integral equation
\begin{equation}\label{integral_relation}
\int_{0}^{1}dy\frac{f_{\rm exp}(y)}{1+uy}=\int_{0}^{\infty}dz \frac{2\,e^{-z}}{1-e^{-2z}}\tanh^2\left(\frac{z}{2\sqrt{1+u}}\right).
\end{equation}
Remarkably, this integral equation can be solved and we show that $f_{\rm exp}(y) = f_{\rm BM}(y)$ obtained for the continuous-time BM in Eq. (\ref{eq:f_bm}). This representation of $f_{\rm BM}(y)$ in Eq. (\ref{integral_relation}) turns out to be useful to compute the moments of $\tau$ explicitly (see \cite{supmat}). We would indeed expect this Brownian scaling form to hold for any jump densities with a finite variance $\sigma^2$, due to the Central Limit Theorem. We have verified numerically this universality for other jump distributions $p(\eta)$ with a finite variance (see Fig. \ref{fig:numeric_rw}). However, for heavy tailed distributions, such as for L\'evy flights with index $0<\mu<2$, we numerically find that, while for large $n$, $P(\tau|n) \approx (1/n) f_{\mu}(\tau/n)$, the scaling function $f_\mu(y)$ depends on $\mu$ (see Fig. \ref{fig:numeric_rw}), except at the endpoints $y = \pm 1$ where it seems that $f_{\mu}(\pm 1) = 1/2$ independently of $0<\mu\leq 2$. The result for $P(\tau = n_{\min}-n_{\max}|n)$ is thus less universal compared to the marginal distributions of $n_{\max}$ and $n_{\min}$ given by the arcsine laws (\ref{levy}).

{\it Brownian bridge.} We now turn to the statistics of $t_{\max}$ and $t_{\min}$ for a BB where the initial $x(0)=0$ and the final positions $x(T)=0$ are identical. The motivation for studying the BB comes from the fact that this will be directly applicable to study KPZ/EW interfaces with periodic boundary conditions in space (see below). For the BB, it is well known \cite{feller50} that $P(t_{\max}|T) = 1/T$, i.e., uniform over $t_{\max} \in [0,T]$. The same result holds for $t_{\min}$ as well. However, we find that the PDF of their difference $\tau = t_{\min} - t_{\max}$ takes the scaling form $P(\tau|T) = (1/T) f_{\rm BB}(\tau/T)$ for all $T$, where the scaling function $f_{\rm BB}(y)$ is nontrivial as in Eq. (\ref{eq:f_bb}). To derive this result, we follow the same path decomposition method as in the case of the BM, except in the third segment $t \in [t_{\min},T]$ where we need to impose that the final position of the trajectory is $x(T) = 0$. Thus, while in the time segments (I) and (II) the probabilities are exactly the same for the BM and the BB, in segment (III) we have simply $G_M(x_{\min},T|\epsilon,t_{\min})$ for the BB (after the shift of the origin to $-x_{\min}$). Taking the product of the three segments, and following the same calculations as in the BM case (see \cite{supmat}), we obtain the result for $f_{\rm BB}(y)$ in Eq.~(\ref{eq:f_bb}). As shown in the SM \cite{supmat}, this result can also alternatively derived by exploiting a mapping, known as Vervaat's construction \cite{vervaat79}, between the BB and the Brownian excursion (the latter being just a BB constrained to stay positive on $[0,T]$). We have also verified these calculations by simulating a BB using an algorithm proposed in  Ref.~\cite{majumdar15} finding excellent agreement (see Fig. \ref{fig:numeric}).     
\par

{\it Fluctuating interfaces.} Our results can be applied directly to $(1+1)$-dimensional KPZ/EW interfaces. Consider a one-dimensional interface growing on a
finite substrate of length $L$, with $H(x,t)$ denoting the height of the interface at position $x$ and time $t$, with $0\leq x \leq L$ \cite{review_kpz1,review_kpz2,spohn_houches}. We study both free and periodic boundary conditions (FBC and PBC respectively). In the former case, the height at the endpoints $x=0$ and $x=L$ are free, while in the latter case $H(0,t) = H(L,t)$. The height field evolves by the KPZ equation~\cite{kpz86}
\begin{equation} \label{eq:kpz}
\frac{\partial H(x,t)}{\partial t}=\frac{\partial^2 H(x,t)}{\partial x^2}+\lambda\left(\frac{\partial H(x,t)}{\partial x}\right)^2+\eta\left(x,t\right),
\end{equation}
where $\lambda \geq 0$ and $\eta(x,t)$ is a Gaussian white noise with zero mean and correlator $\langle\eta\left(x,t\right)\eta\left(x',t'\right)\rangle=2\delta(x-x')\delta(t-t')$. The zero-mode, i.e. the spatially averaged height $\overline{H(t)}=(1/L)\int_{0}^{L}H(x,t)\,dx$, typically grows with time. Hence, the PDF of $H(x,t)$ does not reach a stationary state, even for a finite system. However, the distribution of the relative heights $h(x,t)=H(x,t)-\overline{H(t)}$ does reach a stationary state at long times for finite $L$. For the simpler case of the EW \cite{edwads82} interface ($\lambda=0$ in Eq. (\ref{eq:kpz})), the joint PDF of the stationary relative height $h(x)$, in the case of the FBC, is given by~\cite{majumdar04,comtet05,schehr06}
\begin{equation}\label{eq:stationary}
P_{\rm st}\left(\{h\}\right)=B_L\,e^{-\frac{1}{2}\int_{0}^{L}dx (\partial_{x}h)^2}\delta\left[\int_{0}^{L}h(x)dx\right] \;.
\end{equation}
Here $B_L$ is a normalization constant and the delta-function enforces the zero area constraint $\int_{0}^{L}h(x)dx=0$, which follows from the definition of the relative height. Incidentally, this result for the FBC holds also for the KPZ equation with $\lambda>0$, but only in the large $L$ limit. Thus locally, the stationary height field $h(x)$, for $\lambda \geq 0$, is a BM in space, with $x$ playing the role of time. For the PBC  
an additional factor $\delta(h(0)-h(L))$ is present in Eq. (\ref{eq:stationary}) (see Refs. \cite{majumdar04,comtet05,schehr06}) and moreover it holds for any finite $L$ and arbitrary $\lambda \geq 0$ \cite{supmat}. The PDF of the maximal relative height $h_{\max} = \max_{0 \leq x \leq L} h(x)$ was computed exactly for both boundary conditions \cite{majumdar04,comtet05}.
These distributions are nontrivial due to the presence of the global constraint of zero area under the BM/BB. Under the correspondance $x \Leftrightarrow t$, $L \Leftrightarrow T$ and $h(x) \Leftrightarrow x(t)$, the stationary interface maps onto a BM (for FBC) and to a BB (for PBC) of duration $T$, with an important difference however: the corresponding BM and BB have a zero-area constraint. While this constraint affects the PDF of the maximal (minimal) height, it is clear that the position at which the maximal (respectively minimal) height occurs is not affected due to this global shift by the zero mode. Hence, the distributions of the positions of maximal and minimal height for the stationary KPZ/EW interface is identical to that of $t_{\max}$ and $t_{\min}$ for the BM (for FBC) and of the BB (for PBC). Hence the PDF of $\tau = t_{\min} - t_{\max}$ given in Eqs. (\ref{eq:f_bm}) and (\ref{eq:f_bb}) also gives the PDF of the position-difference between the minimal and maximal heights in the stationary KPZ/EW interface. We have verified this analytical prediction for the KPZ/EW interfaces by solving Eq. (\ref{eq:kpz}) numerically, finding excellent agreement  (see Figs. 4 and 5 in \cite{supmat}).

We have presented an exactly solvable example for the distribution of the time difference between the occurrences of the maximum and the minimum of a stochastic process of a given duration $T$. Our results show that, even for BM or BB, this distribution is highly non-trivial. We have also shown how the same distribution shows up for KPZ/EW interfaces in (1+1) dimensions in their stationary state. Computing this non-trivial distribution for other stochastic processes, such as L\'evy flights, remains a challenging open problem.

%In this Letter, we have derived the exact PDF of $\tau = t_{\min} - t_{\max}$ for a one-dimensional Brownian motion and Brownian bridge of duration $T$. We then applied our results to compute the distribution of the position-difference between the maximum and the minimum height of a fluctuating KPZ/EW interface on a substrate of size $L$, in its stationary state. Furthermore, we also investigated the universality of the Brownian motion result for discrete-time random walks and L\'evy flights. For an $n$-step random walk with a finite jump variance, we have shown that indeed the PDF $P(\tau | n)$ converges, for large $n$, to the Brownian result. 
%
%However, for L\'evy flights of index $0<\mu<2$, we found numerically that $P(\tau | n) \approx (1/n) f_{\mu}(\tau/n)$ for large $n$, where the scaling function $f_\mu(y)$ depends continuously on $\mu$. Calculating this scaling function for general $\mu$ remains a challenging open problem. 

\end{document}

\title{Time between the maximum and the minimum of a stochastic process: supplemental material}% Force line breaks with \\
\author{Francesco Mori$^1$, Satya N. Majumdar$^1$, Gr\'egory Schehr}
\affiliation{ Universit\'e Paris-Sud, CNRS, LPTMS, UMR 8626,
  91405 Orsay, France}

\date{\today}

\begin{abstract}
We give the principal details of the calculations described in the main text of the Letter.
\end{abstract}


\maketitle %\maketitle must follow title, authors, abstract and \pacs

\section{Derivation of the distribution of $\tau = t_{\min} - t_{\max}$ for Brownian motion}

\begin{figure}[h]
  \centering
\includegraphics[width=0.55\linewidth]{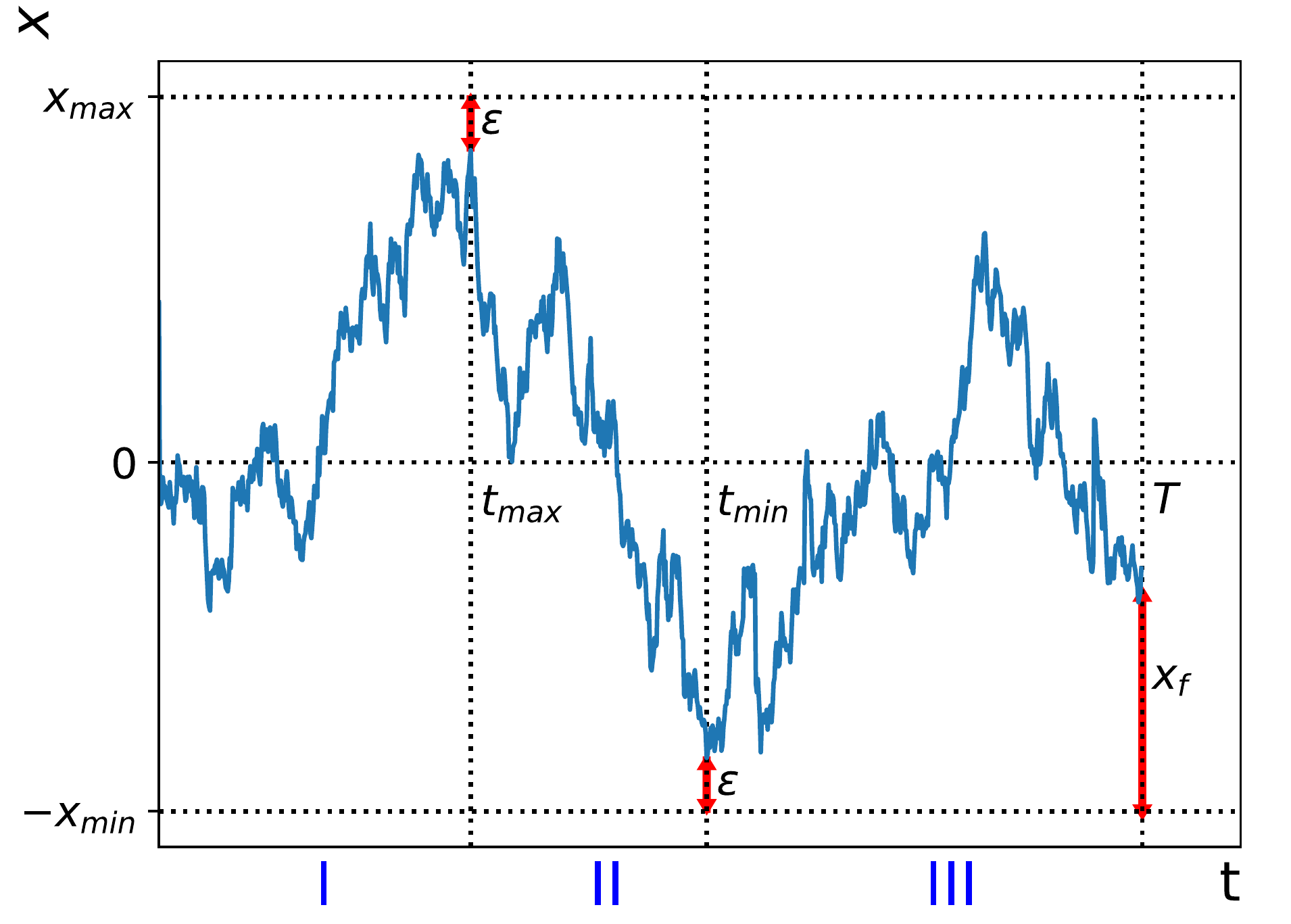}
\caption{A typical trajectory of a Brownian motion $x(t)$ during the time interval $[0,T]$, starting from $x(0)=0$. The value of the global maximum is $x_{\max} - \epsilon$ (with $\epsilon>0$) and the value of the global minimum $-x_{\min} + \epsilon$, where $\epsilon$ is the cut-off needed to enforce absorbing boundary conditions at $x_{\max}$ and $x_{\min}$ (as explained in the text). The time at which the maximum (respectively the minimum) occurs is denoted by $t_{\max}$ (respectively $t_{\min}$). The final position $x(T)$, measured with respect to $-x_{\min}$ is denoted by $x_{\rm f}$. The total time interval $[0,T]$ is divided into three segments: $[0,t_{\max}]$ (I), $[t_{\max}, t_{\min}]$ (II) and $[t_{\min}, T]$ (III), for the case $t_{\min}>t_{\max}$.}
\label{fig:brownian} 
\end{figure}

In this section, we outline the derivation of Eq. (4) of the main text. As explained in the text, the grand joint probability distribution function (PDF) $P(x_{\min},x_{\max},t_{\min},t_{\max}|T)$ can be computed by decomposing the time interval $[0,T]$ into three segments: (I) $[0,t_{\max}]$, (II) $[t_{\max}, t_{\min}]$ and (III) $[t_{\min},T]$. In the first segment (I), the trajectory starts at $0$ at time $t=0$, arrives at $x_{\max}$ at time $t_{\max}$ and stays inside the space interval $x(t) \in [-x_{\min}, x_{\max}]$ during $[0,t_{\max}]$ (see Fig. \ref{fig:brownian}). In the second segment (II), the trajectory starts at $x_{\max}$ at time $t_{\max}$ and arrives at $-x_{\min}$ at time $t_{\min}$ and stays inside the box $[-x_{\min}, x_{\max}]$. Finally, in the third segment (III), the trajectory starts at $-x_{\min}$ at time $t_{\min}$ and stays inside the box $[-x_{\min}, x_{\max}]$ during $[t_{\min},T]$. Clearly the trajectory stays inside the box $[-x_{\min}, x_{\max}]$ because, by definition, it can neither exceed its maximum value $x_{\max}$, nor can it go below its minimum $-x_{\min}$. Thus, to compute the grand joint PDF, we need, as a basic building block, the Green's function $G_M(x,t|x_0,t_0)$ denoting the probability density for a Brownian motion (BM), starting from $x_0$ at $t_0$, to arrive at $x$ at time $t$, white staying inside the box $x(t) \in [0,M]$ during $[t_0,t]$. An explicit expression for this Green's function can be easily computed and is given by \cite{Risken, Redner}
 \begin{equation}\label{eq:g} 
G_M(x,t|x_0,t_0)=\frac{2}{M}\sum_{n=1}^{\infty}\sin\left(\frac{n\pi x}{M}\right)\sin\left(\frac{n \pi x_0}{M}\right)e^{-\frac{n^2\pi^2}{2M^2}(t-t_0)}\;.
\end{equation}

We consider a typical trajectory, as shown in Fig.~\ref{fig:brownian} of the Supplemental Material (SM), where we assume that $t_{\max} < t_{\min}$ (the complementary case $t_{\min}<t_{\max}$ can be easily obtained just by exchanging $t_{\min}$ and $t_{\max}$). As explained in the main text, we have introduced a cut-off $\epsilon$ to regularise the BM such that the actual values of the maximum and the minimum are respectively $x_{\max} - \epsilon$ and $-x_{\min} + \epsilon$ (see Fig. \ref{fig:brownian} of the SM). To make use of the Green's function $G_M$ in Eq. (\ref{eq:g}), we first shift the origin in Fig. \ref{fig:brownian} to $-x_{\min}$ and denote
\begin{eqnarray}\label{def_M}
M = x_{\max} + x_{\min} \geq 0 \;.
\end{eqnarray}

We start with the segment (I), where the probability $P_{\rm I}$ is just proportional to 
\begin{eqnarray} \label{PI_1}
P_{\rm I} \propto G_{M}(M-\epsilon,t_{\max}|x_{\min},0) = \frac{2}{M} \sum_{n_1=1}^\infty \sin \left(\frac{n_1 \pi(M-\epsilon) }{M} \right) \sin \left( \frac{n_1 \pi\, x_{\min}}{M}\right)\, e^{-\frac{n_1^2 \pi^2}{2M^2}\, t_{\max}}\;.
\end{eqnarray}
where we have used the expression (\ref{eq:g}), after shifting the origin to $-x_{\min}$. Expanding to leading order in $\epsilon$, we get
\begin{eqnarray}\label{PI_2}
P_{\rm I} \propto \frac{2 \pi \epsilon}{M^2} \sum_{n_1=1}^\infty (-1)^{n_1-1} n_1 \sin{\left(\frac{n_1 \pi \, x_{\min}}{M} \right)} \,e^{-\frac{n_1^2 \pi^2}{2M^2}\, t_{\max}}\;.
\end{eqnarray}
We next consider the segment (II). Here the probability $P_{\rm II}$ is proportional to 
\begin{eqnarray} \label{PII_1}
P_{\rm II} \propto G_{M}(\epsilon,t_{\min}|M-\epsilon,t_{\max})  \;.
\end{eqnarray} 
Using again Eq. (\ref{eq:g}), and expanding to leading order for small $\epsilon$, we obtain
\begin{eqnarray} \label{PII_2}
P_{\rm II} \propto \frac{2 \pi^2\,\epsilon^2}{M^3} \sum_{n_2=0}^\infty (-1)^{n_2-1} n_2^2 \, e^{-\frac{n_2^2 \pi^2}{2M^2}(t_{\min}-t_{\max})} \;.
\end{eqnarray}
Finally, for the third time segment (III), we obtain similarly
\begin{eqnarray} \label{PIII_1}
P_{\rm III} \propto \int_0^M G_{M}(x_{\rm f},T|\epsilon,t_{\min}) \, dx_{\rm f} \;,
\end{eqnarray}
after integrating over the final position $x_{\rm f} \in [0,M]$.  
Using Eq. (\ref{eq:g}) and expanding for small $\epsilon$ we obtain (after integration over $x_f$)
\begin{eqnarray} \label{PIII_2}
P_{\rm III} \propto \frac{2\epsilon}{M} \sum_{n_3=1}^\infty \left[1 - (-1)^{n_3} \right] \, e^{-\frac{n_3^2 \pi^2}{2M^2} (T-t_{\min})} \;.
\end{eqnarray}
Taking the product of the three segments (\ref{PI_2}), (\ref{PII_2}) and (\ref{PIII_2}), we get the total probability of the trajectory to be proportional to (see also Ref. \cite{schehr10} for a derivation of this formula using real space renormalization group)
\begin{eqnarray}\label{produc_1}
&&P(x_{\min},x_{\max},t_{\min},t_{\max}|T) \propto P_{\rm I} P_{\rm II} P_{\rm III} \propto \frac{\epsilon^4}{M^6} \sum_{n_1=1}^\infty (-1)^{n_1-1} n_1 \sin{\left(\frac{n_1 \pi \, x_{\min}}{M} \right)} \,e^{-\frac{n_1^2 \pi^2}{2M^2}\, t_{\max}}\ \\
&& \times \sum_{n_2=0}^\infty (-1)^{n_2-1} n_2^2 \, e^{-\frac{n_2^2 \pi^2}{2M^2}(t_{\min}-t_{\max})}  \times \sum_{n_3=1}^\infty \left[1 - (-1)^{n_3} \right] \, e^{-\frac{n_3^2 \pi^2}{2M^2} (T-t_{\min})} \;, \quad \quad M = x_{\min} + x_{\max} \geq 0 \nonumber \;.
\end{eqnarray}
By `$\propto$', we have omitted the explicit dependence on the volume factors of the variables, i.e. $dt_{\max}, dt_{\min}, dx_{\max}$ and $dx_{\min}$, since $P(x_{\min},x_{\max},t_{\min},t_{\max}|T)$ is a probability density, and not a probability.

We now want to integrate $x_{\min}$ and $x_{\max}$ over $[0,+\infty)$, in order to obtain the joint PDF $P(t_{\min}, t_{\max}|T)$ for $t_{\min} > t_{\max}$
\begin{eqnarray} \label{Pminmax}
P(t_{\min}, t_{\max}|T) = \int_0^\infty dx_{\min} \, \int_0^\infty dx_{\max} P(x_{\min},x_{\max},t_{\min},t_{\max}|T)  \;.
\end{eqnarray}
We therefore need to evaluate an integral of the type
\begin{eqnarray}\label{Iab}
I(\alpha, \beta) = \int_0^\infty dx_{\min}  \int_0^\infty dx_{\max} \sin \left(\frac{\alpha \, x_{\min}}{x_{\min} + x_{\max}} \right) \frac{1}{(x_{\min} + x_{\max})^6} e^{-\frac{\beta}{(x_{\min} + x_{\max})^2}} \;,
\end{eqnarray}
where $\alpha = n_1 \pi$ and $\beta = \frac{\pi^2}{2}\left(n_1^2 t_{\max} + n_2^2(t_{\min} - t_{\max}) + n_3^2 (T-t_{\min}) \right)$. This integral can be explicitly evaluated by performing the change of variables $M = x_{\min} + x_{\max} $ and $m = x_{\min}$ and we get
\begin{eqnarray} \label{Iab_2}  
I(\alpha, \beta) = \frac{1-\cos \alpha}{2 \alpha \, \beta^2 } \;.
\end{eqnarray} 
Using this result (\ref{Iab_2}) in Eq. (\ref{Pminmax}), one obtains (for $t_{\max} < t_{\min}$)
\begin{eqnarray}\label{Pminmax2}
P_<(t_{\min}, t_{\max}|T)  = A \; \theta(t_{\min} - t_{\max}) \, \sum_{n_1, n_2, n_3 = 1}^\infty \frac{(-1)^{n_2+1} n_2^2  [1-(-1)^{n_1}] [1-(-1)^{n_2}]}{\left[n_1^2 t_{\max} + n_2^2 (t_{\min}-t_{\max})+ n_3^2 (T-t_{\min})\right]^2}  \;,
\end{eqnarray}
where we used $dt_{\max} dt_{\min} \propto \epsilon^4$ and the subscript `$<$' indicates $t_{\max} < t_{\min}$. Here $\theta(x)$ is the Heaviside theta function. In arriving at this final form (\ref{Pminmax2}), we have used that $(-1)^{n_1} = -1$ since only the odd values of $n_1$ contribute to the sum. The overall proportionality constant $A$ has to be fixed from the normalization condition as discussed below. Note that, for the complementary case $t_{\min} < t_{\max}$, one can perform a similar computation, using the product of the three Green's functions, and one obtains 
\begin{eqnarray} \label{symmetry}
P_>(t_{\min},t_{\max}|T) = A \; \theta(t_{\max} - t_{\min})  \sum_{n_1, n_2, n_3 = 1}^\infty (-1)^{n_2+1} n_2^2 \frac{[1-(-1)^{n_1}] [1-(-1)^{n_2}]}{\left[n_1^2 t_{\min} + n_2^2 (t_{\max}-t_{\min})+ n_3^2 (T-t_{\max})\right]^2} \;.
\end{eqnarray}
One sees the symmetry 
\begin{eqnarray} \label{symmetry2}
P_>(t_{\min},t_{\max}|T) = P_<(t_{\max}, t_{\min}|T)  \;.
\end{eqnarray} 
This nontrivial symmetry can be traced back to the fact that the BM is symmetric under reflection $x \to -x$. This constant $A$ can be fixed from the normalisation condition
\begin{eqnarray}\label{normalisation1}
\int_0^T dt_{\min} \int_0^T dt_{\max} \left[P_>(t_{\min}, t_{\max}|T)+  P_<(t_{\min}, t_{\max}|T) \right] = 1 \;.
\end{eqnarray}     
We will see later that the normalisation constant $A$ is given exactly by
\begin{eqnarray} \label{A_1}
A = \frac{4}{\pi^2} \;.
\end{eqnarray}   
   
\vspace*{0.5cm}   
\noindent{\bf Computation of the PDF of $\tau = t_{\min} - t_{\max}$:} To compute the PDF $P(\tau|T)$ of $\tau = t_{\min} - t_{\max}$, we focus on the case $t_{\min}>t_{\max}$, i.e. $\tau >0$. The complementary case $\tau<0$ is simply determined from the symmetry $P(-\tau|T) = P(\tau|T)$, obtained from exchanging $t_{\max}$ and $t_{\min}$ as discussed above. For $\tau >0$, one has
\begin{eqnarray}\label{Ptau_1}
P(\tau|T) = \int_0^T dt_{\max} \int_0^T dt_{\min} P_<(t_{\min}, t_{\max}|T) \; \delta(t_{\min} - t_{\max}-\tau) \;,
\end{eqnarray}
where $P_<(t_{\min}, t_{\max},|T)$ is given in Eq. (\ref{Pminmax2}). Integrating over $t_{\min}$ gives
\begin{eqnarray}\label{Ptau_2}
P(\tau|T) &=& \int_0^{T-\tau} dt_{\max} P_<(t_{\max} + \tau, t_{\max}|T) \\
&=& A \sum_{n_1, n_2, n_3 =1}^\infty (-1)^{n_2+1} n_2^2 (1-(-1)^{n_1})(1-(-1)^{n_2}) \int_0^{T-\tau} \frac{dt_{\max}}{\left( n_1^2 t_{\max} + n_2^2 (t_{\min}-t_{\max})+ n_3^2 (T-t_{\min})\right)^2} \;.\nonumber
\end{eqnarray}
This integral can be explicitly performed, giving
\begin{eqnarray}\label{Ptau_4}
P(\tau|T) &=& A \; (T - \tau)     \sum_{n_1, n_2, n_3 =1}^\infty   (-1)^{n_2+1} n_2^2 \frac{(1-(-1)^{n_1})(1-(-1)^{n_3}) }{(n_1^2 (T-\tau) + n_2^2 \tau)(n_3^2 (T-\tau) + n_2^2 \tau)} \;.
\end{eqnarray}     
Remarkably, the sums over $n_1$ and $n_3$ get decoupled and each yields exactly the same contribution. Hence we get
\begin{eqnarray}\label{PTau_5}
P(\tau |T) = A \; (T-\tau) \sum_{n_2=1}^\infty (-1)^{n_2+1} n_2^2 \left[ \sum_{n=1}^\infty \frac{1-(-1)^n}{n^2(T-\tau)+ n_2^2 \tau}\right]^2 \;.
\end{eqnarray}   
This sum over $n$ inside the parenthesis can be performed using the identity
\begin{eqnarray}\label{identity_Ptau}
\sum_{n=1}^\infty \frac{1-(-1)^ n}{b+ n^2} = \frac{\pi}{2 \sqrt{b}} {\rm tanh}\left( \frac{\pi}{2} \sqrt{b}\right) \;.
\end{eqnarray}  
Using this identity (\ref{identity_Ptau}) into Eq. (\ref{PTau_5}) one obtains, for $\tau >0$, 
\begin{eqnarray}\label{PTau_6}
P(\tau|T) = \frac{1}{T} f_{\rm BM}\left( \frac{\tau}{T}\right) \;, \;\quad {\rm where} \quad \quad f_{\rm BM}(y) = A \; \frac{\pi^2}{4y} \, \sum_{n=1}^\infty (-1)^{n+1} {\rm \tanh}^2 \left( \frac{\pi}{2} n \sqrt{\frac{y}{1-y}}\right) \;.
\end{eqnarray}
Finally, the normalisation constant $A$ is determined from the fact that 
\begin{eqnarray} \label{norm1/2}
\int_0^1 dy f_{\rm BM}(y) = \frac{1}{2} \;.
\end{eqnarray}   
Performing the integral over $y$ and the sum over $n$ explicitly, we get, after a few steps of algebra $A=4/\pi^2$ as in Eq.~(\ref{A_1}). Finally, this gives, using also the symmetry $\tau \to -\tau$, the result 
\begin{eqnarray}\label{eq:f_bm}
f_{\rm BM}(y)=\frac{1}{|y| }\sum_{n=1}^{\infty}(-1)^{n+1}\tanh^2\left(\frac{n\pi}{2}\sqrt{\frac{|y|}{1-|y|}}\right) \;,
\end{eqnarray}
as given in Eq. (4) of the main text. 

\vspace*{0.5cm}
\noindent{\bf Asymptotic analysis of $f_{\rm BM}(y)$.} We consider the function $f_{\rm BM}(y)$ given explicitly in Eq. (\ref{eq:f_bm}) where $y \in [-1,1]$ and $f_{\rm BM}(-y) = f_{\rm BM}(y)$. We first study the limit when $y \to 1$ (or equivalently $y \to -1$). In this limit, we can replace $\tanh^2\left(\frac{n\pi}{2}\sqrt{\frac{|y|}{1-|y|}}\right)$ by $1$, to leading order and evaluate the resulting sum over $n$ as
\begin{eqnarray}\label{yto1}
f_{\rm BM}(y \to 1) = \sum_{n=1}^\infty (-1)^{n+1} \;. 
\end{eqnarray}
Of course, this sum is not convergent. However, one can interpret it in a regularised sense as follows \cite{randon-furling08}
\begin{eqnarray}\label{yto1_2}
f_{\rm BM}(y \to 1) = \lim_{\alpha \to -1} \sum_{n=1}^\infty \alpha^{n+1} = \lim_{\alpha \to -1} \frac{\alpha^2}{1-\alpha} = \frac{1}{2} \;.
\end{eqnarray}

Next, we consider the limit $y \to 0$. In order to investigate this limit, it turns out that the representation given in Eq. (\ref{eq:f_bm}) is not suitable, since the series diverges strongly if one naively takes the limit $y\to 0$. Hence, it is convenient to derive an alternative representation of $f_{\rm BM}(y)$ which will allow us to obtain the $y \to 0$ behaviour correctly. To proceed, we use the Poisson summation formula. Consider the sum
\begin{equation}\label{s_of_a}
s(a)=\sum_{n=0}^{\infty}(-1)^{n+1}\tanh^2(na)=-\sum_{n=0}^{\infty}e^{i \pi n}\tanh^2(na)=-\frac{1}{2}\sum_{n=-\infty}^{\infty}e^{i\pi n}\tanh^2(na) \;.
\end{equation}
This sum can be re-written, using the Poisson summation formula, as
\begin{eqnarray} \label{Poisson1}
s(a) = - \frac{1}{2} \sum_{m=-\infty}^\infty \hat F(2\pi m) \;, \quad \quad {\rm where} \quad \quad \hat F(2\pi m) = \int_{-\infty}^\infty e^{i 2 \pi m x} e^{i \pi x} \tanh^2(a x)\, dx \;.
\end{eqnarray}
This integral can be performed explicitly, using the identity
\begin{eqnarray}\label{Poisson2}
\int_{0}^\infty \cos{(b y)}\,  \tanh^2(y) \, dy = - \frac{\pi}{2} \frac{b}{\sinh\left(\frac{\pi b}{2} \right)} \;.
\end{eqnarray}
Using this, we get 
\begin{eqnarray}\label{Poisson3}
s(a) = \frac{\pi^2}{2a^2} \sum_{m=-\infty}^{\infty}\frac{2m+1}{\sinh\left(\frac{\pi^2 (2m+1)}{2a}\right)}.
\end{eqnarray}
Using $a = \frac{\pi}{2} \sqrt{|y|/(1-|y|)}$ in Eq. (\ref{eq:f_bm}) and Eq. (\ref{Poisson3}), we obtain an exact alternative representation of $f_{\rm BM}(y)$ as
\begin{eqnarray}\label{Poisson4}
f_{\rm BM}(y) =\frac{1}{|y| }\sum_{n=1}^{\infty}(-1)^{n+1}\tanh^2\left(\frac{n\pi}{2}\sqrt{\frac{|y|}{1-|y|}}\right) =  \frac{2(1-|y|)}{|y|} \sum_{m = -\infty}^\infty \frac{2m+1}{\sinh \left( (2m+1) \pi \sqrt{\frac{1-|y|}{|y|}} \right)}  \;.
\end{eqnarray}
One can now take the $y \to 0$ limit in the last expression, where the terms $m=0$ and $m=-1$ dominate in this limit. This yields, to leading order,  
\begin{equation}\label{asympt_y0}
f_{\rm BM}(y) \approx \frac{8}{y^2}e^{-{\pi}/{\sqrt{y}}} \;, \quad {\rm as} \quad y \to 0 \;.
\end{equation}
Hence the asymptotic behaviors of $f_{\rm BM}(y)$ can be summarised as
\begin{eqnarray}\label{summary_asymptotics}
f_{\rm BM}(y) \approx
\begin{cases}
&\dfrac{1}{2} \quad \; \quad \quad \quad \; {\rm as} \quad y \to 1 \\
& \\
& \dfrac{8}{y^2}\, e^{-{\pi}/{\sqrt{y}}}   \quad {\rm as} \quad y \to 0 
\end{cases}
\end{eqnarray}

\section{The distribution of $\tau = t_{\min} - t_{\max}$ for discrete-time random walks: exact result for the double-exponential jump distribution}\label{sec:exponential}

In the previous section, we have provided the derivation of the exact formula for the distribution of the time difference $\tau = t_{\min} - t_{\max}$ between the minimum and the maximum of a BM which, of course, takes place in continuous time. One interesting question concerns the universality of this distribution. For example, instead of the continuous-time BM, if we consider random walks (RW) in discrete time but in a continuous space, would this distribution still hold asymptotically for a large number of steps? More precisely, let us consider a walker on a line whose position after $k$ steps evolves as
\begin{equation} \label{def_rw_sm}
x_k=x_{k-1}+\eta_k \;,
\end{equation}
starting from $x_0 = 0$ and where $\eta_k$'s are the jumps at different time steps, each chosen independently from a common distribution $p(\eta)$, with zero mean. Note that this model (\ref{def_rw_sm}) includes also L\'evy flights, for which $p(\eta)$ has a power law tail $p(\eta) \sim 1/|\eta|^{1 +\mu}$ for $|\eta| \gg 1$, with the L\'evy index $0 < \mu \leq 2$. 

Let us consider our walk up to $n$ steps and denote by $n_{\min}$ and $n_{\max}$ the step numbers at which the minimum $x_{\min}$ and the maximum $x_{\max}$ occur respectively. Clearly, $n_{\min}$ and $n_{\max}$ are random variables and here we are interested in the distribution of the time difference $\tau = n_{\min} - n_{\max}$, for arbitrary symmetric and continuous jump distribution $p(\eta)$. If the variance $\sigma^2 = \int \eta^2\, p(\eta) d\eta$ is finite, the central limit theorem (CLT) predicts that, for large $n$, the discrete-time process $x_n$ converges to the continuous-time Brownian motion and, hence, one would expect that, for finite $\sigma^2$, the distribution $P(\tau|n)$, for large $n$, should also converge to the Brownian motion result, namely
\begin{eqnarray} \label{asympt_Ptau}
P(\tau|n) \underset{n \to \infty}{\longrightarrow} \frac{1}{n} f_{\rm BM}\left( \frac{\tau}{n}\right) \;,
\end{eqnarray}
where the scaling function $f_{\rm BM}(y)$ is given in Eq. (\ref{eq:f_bm}). Note that the validity of this result requires a finite $\sigma^2$, but the scaling function $f_{\rm BM}(y)$ is independent of the actual value of $\sigma^2$. It would be nice to verify this expectation, based on the CLT, by computing exactly $P(\tau|n)$ in some solvable cases of discrete-time RW (\ref{def_rw_sm}). Below, we provide the exact solution for the special case of the double exponential jump distribution
\begin{eqnarray}\label{exp_PDF}
p(\eta) = \frac{1}{2} \, e^{-|\eta|} \;.
\end{eqnarray}
In this case $\sigma^2 = 2$ is finite and our exact solution indeed verifies that the CLT prediction in Eq. (\ref{asympt_Ptau}) is correct. For other jump distributions, it turns out that the exact computation of $P(\tau|n)$ is very hard. We have verified numerically that, for a class of jump distributions with finite $\sigma^2$, indeed the CLT prediction holds and we get the scaling function $f_{\rm BM}(y)$ (see Fig. 3 in the text). However, for L\'evy flights with index $0 < \mu < 2$, for which $\sigma^2$ is divergent, we find numerically that, for large $n$, the following scaling holds 
\begin{eqnarray} \label{asympt_Ptau_mu}
P(\tau|n) \underset{n \to \infty}{\longrightarrow} \frac{1}{n} f_{\mu}\left( \frac{\tau}{n}\right) \;,
\end{eqnarray}
but the scaling function $f_{\mu}(y)$ depends explicitly on $\mu$ and differs from the Brownian result $f_{\rm BM}(y)$ (see Fig. 3 in the text). Note however that, at the endpoints, $f_\mu(\pm 1) = 1/2$ seems to be universal numerically, i.e., independent of $\mu$, although the full scaling function $f_\mu(y)$ does depend on $\mu$. 

The computation of $f_{\mu}(y)$ remains an outstanding open problem. 

\begin{figure}
  \centering
\includegraphics[width = 0.6 \linewidth]{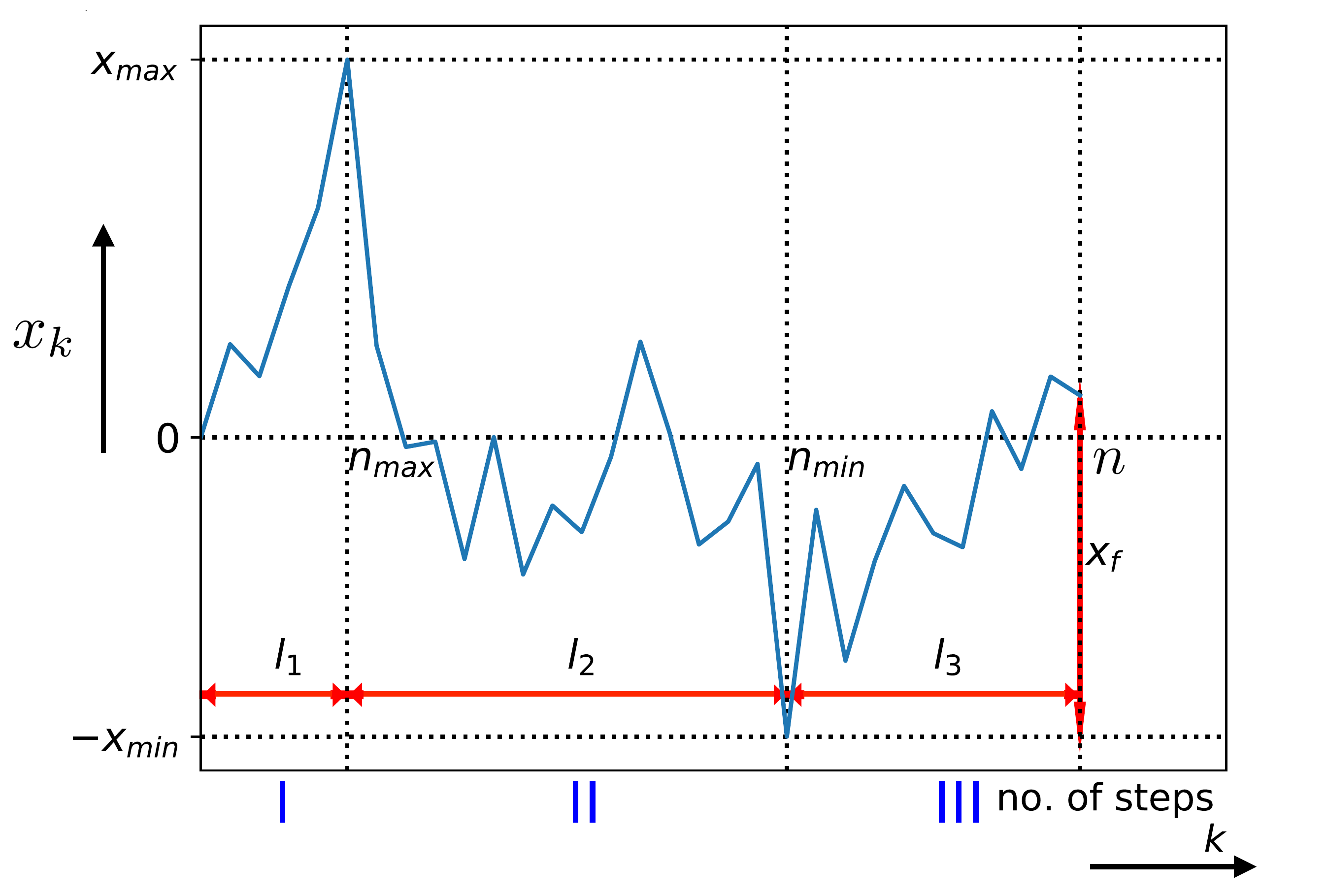}
\caption{A typical trajectory of a discrete-time random walk $x_k$ versus $k$ up to $n$ steps, starting from $x_0=0$. The global maximum $x_{\max}$ occurs at step $n_{\max}$ and the global minimum $-x_{\min} \leq 0$ at step $n_{\min}$. For this trajectory $n_{\min} > n_{\max}$. The final position of the walker at step $n$ is denoted by $x_{\rm f}$ measured with respect to the global minimum $-x_{\min}$. The total duration of $n$ steps has been divided into three segments: $0\leq k \leq n_{\max}$ (I), $n_{\max}\leq k \leq n_{\min}$ (II) and $n_{\min} \leq k \leq n$ (III). The respective durations of these three segments are denoted by $l_1$, $l_2$ and $l_3$.}
\label{fig:exponential} 
\end{figure}

\vspace*{0.5cm}
\noindent {\bf Exact solution for the double exponential jump distribution.} Consider a typical trajectory of $n$ steps of the discrete RW described in Eq. (\ref{def_rw_sm}) with $\eta_k$'s drawn independently from a symmetric and continuous distribution $p(\eta)$, see Fig. \ref{fig:exponential} in the SM. For the moment, we consider an arbitrary jump distribution $p(\eta)$ and later, we will focus on the specific case of a double exponential distribution as in Eq. (\ref{def_rw_sm}). As in the BM case, our strategy will be to first compute the grand joint PDF $P(x_{\min}, x_{\max}, n_{\min}, n_{\max} | n)$ of the four random variables $x_{\min}, x_{\max}, n_{\min}$ and $n_{\max}$ and then integrate out $x_{\min}$ and $x_{\max}$ to obtain the joint distribution $P(n_{\min}, n_{\max} | n)$. To compute this grand PDF, we again divide the interval $[0,n]$ into three segments of lengths $l_1=n_{\max}$, $l_2=n_{\min}-n_{\max}$ and $l_3=n-n_{\min}$ (see Fig. \ref{fig:exponential} in the SM) -- here again we consider the case $n_{\max} < n_{\min}$ (the complementary case can be solved as in the Brownian case). The grand PDF can then be written as the product of 
the probabilities $P_{\rm I}$, $P_{\rm II}$ and $P_{\rm III}$ of the three independent segments: $0\leq k \leq n_{\max}$ (${\rm I}$), $n_{\max}\leq k \leq n_{\min}$ (${\rm II}$) and $n_{\min} \leq k \leq n=l_1+l_2+l_3$ (${\rm III}$). These probabilities can be written in terms of the basic building block, namely the restricted Green's function $G\left(x,n | M \right)$ defined as the probability density to reach $x$ in $n$ steps starting from the origin $x_0=0$ while staying inside the box $[0,M]$, i.e., with the constraint that $x_k\in[0,M]$ for all $k=1, \ldots, n$, 
\begin{eqnarray}\label{def_restricted}
G\left(x,n | M \right) = {\rm Prob.} \left[x_0=0 \, , \, 0 \leq x_1 \leq M \,, \, 0 \leq x_2 \leq M \,, \, \ldots,  \, 0 \leq x_{n-1} \leq M \,, \, 0 \leq x_n = x \leq M \right] \;.
\end{eqnarray}

To proceed, we first notice that the probability of each segment can be expressed in terms of the restricted Green's function $G\left(x,n| M \right)$ as follows. In order to do this, it is crucial to use that the jump distribution $p(\eta)$ is symmetric which makes the walk reflexion symmetric around the origin. This is best explained with the help of the Fig. \ref{fig:exponential}. Let us first set $M = x_{\min} + x_{\max}$. In segment I, the trajectory has to start at the origin and reach the level $x_{\max}$ at step $l_1$, while staying in the box $[-x_{\min},x_{\max}]$. Using the invariance under the reflection $x \to -x$, followed by a shift of the origin to the level $x_{\max}$, this probability is just 
\begin{eqnarray}\label{PI}
P_{\rm I}=G\left(x_{\max},l_1 | M\right) \;.
\end{eqnarray}
For the second segment, the trajectory starts at $x_{\max}$ and ends at $- x_{\min}$ at step $l_2$, while staying inside the box $[-x_{\min}, x_{\max}]$ (see Fig. \ref{fig:exponential}). Using a similar argument as in the previous case, one gets 
\begin{eqnarray}\label{PII}
P_{\rm II}=G\left(M,l_2 | M \right) \;.
\end{eqnarray}
For the third segment, the trajectory starts at $-x_{\min}$ and stays inside the box $[-x_{\min}, x_{\max}]$ up to $l_3$ steps. Thus this probability is given by  
\begin{eqnarray}\label{PIII}
P_{\rm III}=\int_0^M G\left(x_{\rm f},l_3 | M \right) dx_{\rm f} \;,
\end{eqnarray}
where $x_{\rm f}$ is the final position of the trajectory measured with respect to $-x_{\min}$ (see Fig. \ref{fig:exponential}), and we integrated over the final position $x_{\rm f}$. The grand joint PDF $P(x_{\min}, x_{\max}, n_{\min}, n_{\max} | n)$ is given by the product of the three segments 
\begin{eqnarray}\label{product}
P(x_{\min}, x_{\max}, n_{\min}, n_{\max} | n) = P_{\rm I} \, P_{\rm II} \, P_{\rm III} = G\left(x_{\max},l_1 | M\right)\, G\left(M,l_2 | M \right) \, \int_0^M G\left(x_{\rm f},l_3 | M \right) dx_{\rm f}  \;,
\end{eqnarray}
where we recall that 
\begin{eqnarray}\label{def_l}
l_1 = n_{\max} \quad, \quad l_2 = n_{\min} - n_{\max} \quad, \quad l_3 = n-n_{\min} \quad {\rm and} \quad M = x_{\max} + x_{\min} \;.
\end{eqnarray}
Note that, unlike the BM, for the discrete-time RW, we do not need to put a cut-off $\epsilon$. 

Thus, from Eq. (\ref{product}), we see that the main ingredient needed to compute the grand PDF 
is the restricted Green's function $G\left(x,n|M\right)$ defined in Eq. (\ref{def_restricted}). Using the Markov property of the
process with an arbitrary jump distribution $p(\eta)$, we can easily write down a
recursion relation for $G\left(x,n|M\right)$
\begin{equation}
\label{recursive}
G\left(x,n | M \right)=\int_{0}^{M} dx' G\left(x',n-1|M \right)p(\eta = x-x') \;,
\end{equation}
valid for all $n \geq 1$ and starting from the initial condition $G(x,0|M) = \delta(x)$. This equation can be understood as follows. Let the walk arrive at $x' \in [0,M]$ at step $n-1$ (without leaving the box $[0,M]$ up to $n-1$ steps) and then it makes a jump from $x'$ to $x$ at the $n^{\rm th}$ step. 
The probability for this jump is simply $p(\eta = x-x')$. Remarkably, this simple equation (\ref{recursive}) can not be solved
exactly for arbitrary jump distribution $p(\eta)$. The reason is because the limits of the integral are over a finite range $[0,M]$.
In the semi-infinite case $M \to \infty$, this reduces to Wiener-Hopf equation which can be solved for arbitrary symmetric and continuous
$p(\eta)$. Although the solution in this case is not fully explicit for $G\left(x,n | M \to \infty \right)$, one can obtain an explicit expression for
its generating function in terms of the Fourier transform of the jump distribution $p(\eta)$. This is known as the Ivanov formula \cite{ivanov} (see also the Appendix A of Ref. \cite{mounaix} for a transparent derivation). Unfortunately, for finite $M$, no exact solution is known for arbitrary $p(\eta)$. However, for the double exponential jump distribution in Eq. (\ref{exp_PDF}), we can obtain an exact solution of Eq. (\ref{recursive}) for finite $n$, as shown below.

To proceed, we first consider the generating function 
\begin{equation} \label{def_GF}
\tilde{G}\left(x,s | M \right)=\sum_{n=1}^{\infty}G\left(x,n | M \right)s^n \;.
\end{equation}
By multiplying Eq. (\ref{recursive}) by $s^n$, summing over $n$ and using the initial condition $G(x,0|M) = \delta(x)$, we get
\begin{equation}\label{g_tilde}
\tilde{G}\left(x,s|M\right)=s\int_{0}^{M} dx' \tilde{G}\left(x',s|M \right)\, p(x-x')+s\,p(x).
\end{equation}
The double-exponential distribution in Eq. (\ref{exp_PDF}) has the special property that if we differentiate twice, it satisfies a
simple differential equation  
\begin{equation} \label{exp_relation}
p''(x)=p(x)-\delta(x) \;.
\end{equation}
Using this relation, we can then reduce the integral equation in (\ref{g_tilde}) into a differential equation, which then is easier to solve. Differentiating twice Eq. (\ref{g_tilde}) with respect to $x$, and using the identity (\ref{exp_relation}), we get
\begin{equation} \label{differential}
\frac{\partial^2 \tilde{G}\left(x,s|M \right) }{\partial x^2}= (1-s)\tilde{G}\left(x,s|M \right)-s\,\delta(x),
\end{equation}
for $0\leq x\leq M$. For $x>0$, the $\delta$-function in (\ref{differential}) disappears and the general 
solution reads simply
\begin{equation} \label{solution_general}
\tilde{G}\left(x,s|M\right)=A(s,M)\,e^{-\sqrt{1-s}\,x}+B(s,M)\,e^{\sqrt{1-s}\,x} \;,
\end{equation}
where $A(s,M)$ and $B(s,M)$ are two arbitrary constants. In going from the integral (\ref{g_tilde}) to the differential (\ref{differential}) equation, 
we have taken derivatives and hence one has to ensure that the solution of the differential equation also satisfies the integral equation. This
will be true only for specific values of $A(s,M)$ and $B(s,M)$ which then fixes these unknown constants. Indeed, by substituting Eq. (\ref{solution_general}) 
into the integral equation (\ref{g_tilde}), it is straightforward to check that 
\begin{eqnarray}
&& A(s,M)=\frac{1-\sqrt{1-s}}{1-\left(\frac{1-\sqrt{1-s}}{1+\sqrt{1-s}}\right)^2 \, e^{-2\sqrt{1-s} \,M}},  \label{A} \\
&& B(s,M)=-A(s,M)\frac{1-\sqrt{1-s}}{1+\sqrt{1-s}} \, e^{-2\sqrt{1-s}\,M} \;. \label{B}
\end{eqnarray}
Hence, the final exact solution reads
\begin{equation}\label{g_solution}
\begin{split}
\tilde{G}\left(x,s|M \right)=A(s,M)\left[e^{-\sqrt{1-s}\,x}-\frac{1-\sqrt{1-s}}{1+\sqrt{1-s}}e^{-\sqrt{1-s}\,(2M-x)}\right] \;,
\end{split}
\end{equation}
with the amplitude $A(s,M)$ given in Eq. (\ref{A}). 

In order to use this solution (\ref{g_solution}) in Eq. (\ref{product}), we need to take the generating function of the grand PDF $P(x_{\min}, x_{\max}, n_{\min}, n_{\max} | n)$. It is convenient to express this grand PDF in terms of the intervals $l_1, l_2$ and $l_3$. Hence, we write 
\begin{eqnarray}\label{PDF_l1l2l3}
P(x_{\min}, x_{\max}, n_{\min}, n_{\max} | n) \equiv P(x_{\min}, x_{\max}, l_1, l_2, l_3) \;,
\end{eqnarray}  
where $l_1, l_2$ and $l_3$ are given in Eq. (\ref{def_l}). We now multiply Eq. (\ref{product}) by $s_1^{l_1}s_2^{l_2}s_3^{l_3}$ and sum over $l_1$, $l_2$ and $l_3$ to obtain
\begin{equation}\label{laplace}
\sum_{l_1,l_2,l_3=1}^{\infty}
P\left(x_{\min},x_{\max},l_1,l_2,l_3\right)s_1^{l_1}s_2^{l_2}s_3^{l_3}=\tilde{G}\left(x_{\max},s_1|M \right)\tilde{G}\left(M,s_2|M \right)\, \int_0^M dx_{\rm f} \, \tilde{G}\left(x_{\rm f},s_3|M \right) \;,
\end{equation}
where we recall that $M = x_{\min} + x_{\max}$. Using the solution in Eq. (\ref{g_solution}), the integral in the third term in Eq. (\ref{laplace}) can be performed explicitly, giving
\begin{equation}
I(M,s_3)=\int_{0}^{M}dx_{\rm f} \, \tilde{G}\left(x_{\rm f},s_3|M\right)=\frac{A(s_3,M)}{\sqrt{1-s_3}}\left[1-\frac{2}{1+\sqrt{1-s_3}}
e^{-\sqrt{1-s_3}M}+\frac{1-\sqrt{1-s_3}}{1+\sqrt{1-s_3}}e^{-2\sqrt{1-s_3}M}\right].
\end{equation}
To obtain the marginal joint distribution of $n_{\min}$ and $n_{\max}$, we still need to integrate over $x_{\min}$ and $x_{\max}$ in Eq. (\ref{laplace}). Let us first define 
\begin{eqnarray}\label{Pl1l2l3}
P(l_1, l_2,l_3) = \int_{0}^\infty dx_{\min} \, \int_{0}^\infty dx_{\max} \,  P\left(x_{\min},x_{\max},l_1,l_2,l_3\right) \;.
\end{eqnarray}
We can perform this double integral by making a change of variables $(x_{\min}, x_{\max}) \to (x_{\min}, M = x_{\max} + x_{\min})$. Using the explicit expression of $\tilde G$ from Eq. (\ref{g_solution}), and performing the double integral yields
\begin{equation}
\sum_{l_1,l_2,l_3=1}^{\infty}
P\left(l_1,l_2,l_3\right)s_1^{l_1}s_2^{l_2}s_3^{l_3}= \int_{0}^{\infty}dM\int_{0}^{M}dx_{\min} \tilde{G}\left(M-x_{\min},s_1|M \right)\tilde{G}\left(M,s_2|M \right)I(M,s_3).
\end{equation}
Thus, we find
\begin{equation} \label{ugly1}
\begin{split}
\sum_{l_1,l_2,l_3=1}^{\infty}
P\left(l_1,l_2,l_3\right)s_1^{l_1}s_2^{l_2}s_3^{l_3}= \int_{0}^{\infty}dM  
\frac{A(s_1,M)e^{-\sqrt{1-s_1}M}}{\sqrt{1-s_1}}\left[e^{\sqrt{1-s_1}M}+\frac{1-\sqrt{1-s_1}}{1+\sqrt{1-s_1}}e^{-\sqrt{1-s_1}M}-\frac{2}{1+\sqrt{1-s_1}}\right] \\ \times
A(s_2,M)\left[e^{-\sqrt{1-s_2}M}-\frac{1-\sqrt{1-s_2}}{1+\sqrt{1-s_2}}e^{-\sqrt{1-s_2}M}\right]  \frac{A(s_3,M)}{\sqrt{1-s_3}}\left[1-\frac{2}{1+\sqrt{1-s_3}}
e^{-\sqrt{1-s_3}M}  + \frac{1-\sqrt{1-s_3}}{1+\sqrt{1-s_3}}e^{-2\sqrt{1-s_3}M}\right].
\end{split}
\end{equation}
We now use the expression of $A(s,M)$ from Eq. (\ref{A}) and carry out the integral. This rather cumbersome expression can be written
in a slightly more compact form by introducing the variables $\omega_i=\sqrt{1-s_i}$ for $i=1,2,3$. In terms of these variables, Eq. (\ref{ugly1}) reads
\begin{eqnarray}\label{l1_l2_l3}
&&\sum_{l_1,l_2,l_3=1}^{\infty}
P\left(l_1,l_2,l_3\right)s_1^{l_1}s_2^{l_2}s_3^{l_3} \nonumber \\
%&&=\frac{2\omega_2(1-\omega_1)(1-\omega_2)(1-\omega_3)}{\omega_1(1+\omega_2)\omega_3}\int_{0}^{\infty}dM
%\frac{e^{-\omega_2 M}}{1+\frac{1-\omega_1}{1+\omega_1}e^{-\omega_1 M}} 	\left( 1-e^{-\omega_1 M} \right) \\ \times
%\frac{1}{1+\frac{1-\omega_3}{1+\omega_3}e^{-\omega_3 M}} 	\left( 1-e^{-\omega_3 M} \right)
%\frac{1}{1+\left(\frac{1-\omega_2}{1+\omega_2}\right)^2 e^{-2 \omega_2 M}} \;.
&&=\frac{2\omega_2(1-\omega_1)(1-\omega_2)(1-\omega_3)}{\omega_1(1+\omega_2)\omega_3} \int_0^\infty dM \frac{e^{-\omega_2 M}(1-e^{-\omega_1 M})(1-e^{-\omega_3 M})}{\left(1 + \frac{1-\omega_1}{1+\omega_1}\, e^{-\omega_1 M}\right) \left(1 - \left(\frac{1-\omega_2}{1+\omega_2}\right)^2\, e^{-2\omega_2 M}\right) \left(1 + \frac{1-\omega_3}{1+\omega_3}\, e^{-\omega_3 M}\right)} \;.
\end{eqnarray}

We want to compute the PDF $P(\tau|n)$ of $\tau = n_{\min} - n_{\max}$, for a given total number of steps $n$. We can express $P(\tau|n)$ in terms of the joint PDF $P\left(l_1,l_2,l_3\right)$ computed above as follows
\begin{eqnarray}\label{relation1}
P(\tau|n) = \sum_{l_1, l_3=1}^\infty P(l_1, l_2 = \tau, l_3) \, \delta(l_1+l_2+l_3-n) \;.
\end{eqnarray}
Taking the double generating function of this expression (\ref{relation1}) gives
\begin{eqnarray}\label{relation2}
\sum_{\tau, n=1}^\infty P(\tau|n) s_2^\tau \, s^n = \sum_{l_1,\tau,l_3} P(l_1, l_2 = \tau, l_3) \, s^{l_1} (s\, s_2)^{\tau}\, s^{l_3} \;.
\end{eqnarray}
Notice that the right hand side of Eq. (\ref{relation2}) can be read off Eq. (\ref{l1_l2_l3}) by setting $s_1 \to s$, $s_2 \to s\, s_2$ and $s_3 \to s$.  By defining further $\omega=\sqrt{1-s}$ and $\tilde{\omega}=\sqrt{1-ss_2}$, we get
\begin{equation}\label{relation3}
\begin{split}
\sum_{\tau,n = 1}^\infty P\left(\tau |n\right)\,s_2^{\tau}\,s^n=
\frac{2\tilde{\omega}(1-\tilde{\omega})(1-\omega)^2}{(1+\tilde{\omega})\omega^2}\int_{0}^{\infty}dM  
\frac{e^{-\tilde{\omega} M} \left( 1-e^{-\omega M} \right)^2     }{\left[1+\frac{1-\omega}{1+\omega}e^{-\omega M}\right]^2  \left[ 1-\left(\frac{1-\tilde{\omega}}{1+\tilde{\omega}}\right)^2 e^{-2 \tilde{\omega} M}\right]   } \;.
\end{split}
\end{equation}
This expression is exact and from it, we want to extract the asymptotic behavior of $P\left(\tau |n\right)$ for large $n$. 

\vspace*{0.5cm}
\noindent{\bf Large $n$ asymptotics of $P(\tau | n)$.} In this limit, we expect that $P(\tau | n)$ should approach a scaling form 
\begin{eqnarray}\label{scaling_form}
P(\tau | n) \underset{n \to \infty}{\longrightarrow} \frac{1}{n} f_{\rm exp} \left( \frac{\tau}{n}\right) \;.
\end{eqnarray}
Our goal now is to extract this scaling function $f_{\rm exp}(y)$ from the exact formula (\ref{relation3}) and show that $f_{\rm exp}(y) = f_{\rm BM}(y)$ given in Eq. (\ref{Poisson4}). Since we are interested in the scaling limit $\tau, n \to \infty$ keeping the ratio $y = \tau/n$ fixed, we need to investigate the generating function in Eq. (\ref{relation3}) also in the corresponding scaling limit. It is convenient to first parametrize the Laplace variables as $s=e^{-\lambda}$ and  $s_2=e^{-\lambda_2}$. In these new variables, the scaling limit corresponds to  
$\lambda,\lambda_2\rightarrow 0 $ with $\lambda_2/\lambda$ fixed. In this limit, the double sum in the left hand side of Eq. (\ref{relation3}) can be replaced by a double integral. Taking the limit $\lambda,\lambda_2\rightarrow 0$ keeping the ratio $\lambda_2/\lambda$ fixed on both sides of Eq. (\ref{relation3}), we get 
\begin{equation} \label{scaling1}
\int_{0}^{\infty}dn \int_{0}^{n}d \tau \, P\left(\tau | n \right)\,e^{-\lambda_2 \tau}e^{-\lambda \, n} \approx
\frac{2\sqrt{\lambda+\lambda_2}}{\lambda}\int_{0}^{\infty}dM  
\frac{e^{-\sqrt{\lambda+\lambda_2} M}\left(1-e^{-\sqrt{\lambda}M}\right)^2}{\left(1+e^{-\sqrt{\lambda}M}\right)^2\left(1-e^{-2\sqrt{\lambda+\lambda_2}M}\right)}.
\end{equation}
Rescaling $z=\sqrt{\lambda+\lambda_2}\,M$ in the integral on the right hand side leads to
\begin{equation}\label{integral1}
\int_{0}^{\infty}dn\int_{0}^{n}d \tau P\left(\tau | n\right)e^{-\lambda_2 \tau}e^{-\lambda n}=
\frac{2}{\lambda}\int_{0}^{\infty}dz  
\frac{e^{-z}\left(1-e^{-\sqrt{\frac{\lambda}{\lambda+\lambda_2}}z}\right)^2}{\left(1+e^{-\sqrt{\frac{\lambda}{\lambda+\lambda_2}}z}\right)^2\left(1-e^{-2z}\right)}.
\end{equation}
Substituting the scaling form (\ref{scaling_form}) on the left hand side of Eq. (\ref{integral1}) gives 
\begin{equation} \label{integral2}
\begin{split}
\int_{0}^{\infty}dn\int_{0}^{n}d \tau \frac{1}{n}f_{\rm exp} \left(\frac{\tau}{n}\right)\, e^{-\lambda_2 \tau}e^{-\lambda n}=
\int_{0}^{\infty}dn\int_{0}^{1}dy\, f_{\rm exp}(y)\,e^{-\lambda_2 y}e^{-\lambda n}=\int_{0}^{1}dy\frac{f_{\rm exp}(y)}{\lambda+\lambda_2 y}=\frac{1}{\lambda}\int_{0}^{1}dy\frac{f_{\rm exp}(y)}{1+\frac{\lambda_2}{\lambda} y}.
\end{split}
\end{equation}
Comparing this left hand side (\ref{integral2}) with the right hand side of Eq. (\ref{integral1}), with $u=\frac{\lambda_2}{\lambda}$ fixed, we get the identity
\begin{equation}\label{integral2}
\int_{0}^{1}dy\frac{f_{\exp}(y)}{1+uy}=2\int_{0}^{\infty}dz \frac{e^{-z}}{1-e^{-2z}}\tanh^2\left(\frac{z}{2\sqrt{1+u}}\right),
\end{equation}
which is precisely Eq. (10) in the main text. 

\vspace*{0.5cm}
\noindent{\bf Computation of the scaling function $f_{\rm exp}(y)$.} The next step is to invert this integral equation (\ref{integral2}) to obtain $f_{\rm exp}(y)$ explicitly. For this, it is convenient to first rewrite Eq. (\ref{integral2}) in terms of the variables $u=-\frac{1}{w}$ on the left hand side and $t=\frac{z}{2\sqrt{1+u}}$ on the right hand side. This gives
\begin{equation}\label{stieltjes}
\int_{0}^{1}dy\frac{f_{\rm exp}(y)}{w-y}=\frac{2}{w}\sqrt{1-\frac{1}{w}}\int_{0}^{\infty}dt \frac{\tanh^2(t)}{\sinh \left(2t\sqrt{1-\frac{1}{w}}\right)} \;.
\end{equation}
We now recognise the left hand side of Eq. (\ref{stieltjes}) as the Stieltjes transform of the function $f_{\rm exp}(y)$. This Stieltjes transform of this type can be inverted using the so-called Sochocki-Plemelj formula (see for instance the book \cite{mushk_book}). Setting $w = y + i \epsilon$ with $y$ real, this formula reads in our case
\begin{equation}\label{antitransform}
f_{\rm exp}(y)=-\frac{1}{\pi}\lim_{\epsilon\rightarrow 0}\operatorname{Im}\left[  \frac{2}{(y+i\epsilon)}\sqrt{1-\frac{1}{(y+i\epsilon)}}\int_{0}^{\infty}dt \frac{\tanh^2(t)}{\sinh\left(2t\sqrt{1-\frac{1}{(y+i\epsilon)}}\right)} \right].
\end{equation}
We first expand the integrand of the right hand side of Eq. (\ref{antitransform}) for small $\epsilon$ and take the imaginary part
\begin{equation} \label{antitransform2}
\operatorname{Im}\left[\frac{1}{(y+i\epsilon)}\sqrt{1-\frac{1}{(y+i\epsilon)}}\frac{1}{\sinh\left(2t\sqrt{1-\frac{1}{(y+i\epsilon)}}\right)}\right]\simeq
\frac{\epsilon}{y^3}\frac{t \cos\left(2t\sqrt{\frac{1-y}{y}}\right)-\frac{3-2y}{2}\sqrt{\frac{y}{1-y}}\sin\left(2t\sqrt{\frac{1-y}{y}}\right)}{\sin^2\left(2t\sqrt{\frac{1-y}{y}}\right)+\frac{\epsilon^2t^2}{y^3(1-y)}\cos^2\left(2t\sqrt{\frac{1-y}{y}}\right)}.
\end{equation}
Note that we have kept the leading term of order ${\cal O}(\epsilon^2)$ in the denominator on the right hand side in Eq. (\ref{antitransform2}), in order that the integral over $t$ does not diverge. Substituting Eq. (\ref{antitransform2}) in Eq. (\ref{antitransform}) and making the change of variable $v=2t\sqrt{\frac{1-y}{y}}$, we get
\begin{equation}\label{antitransform3}
\begin{split}
f_{\rm exp}(y)= \lim_{\epsilon \to 0}\left[-\frac{\epsilon}{2\pi y^2(1-y)}\int_{0}^{\infty}dv \tanh^2\left(\frac{v}{2}\sqrt{\frac{y}{1-y}}\right)\frac{v\cos(v)-(3-2y)\sin(v)}{\sin^2(v)+\frac{\epsilon^2 v^2}{\left(2y\left(1-y\right)\right)^2} \cos^2(y)} \right] \;.
\end{split}
\end{equation}
To compute the integral in the right hand side, we split it as a sum of integrals over $v \in [0,\pi/2]$ and $ v \in [n\pi-\pi/2,n\pi+\pi/2]$ for $n\geq1$. The integral over $[0,\pi/2]$ is convergent (since there is no divergence of the integrand even when $\epsilon \to 0$ in the denominator) and is of order ${\cal O}(\epsilon)$. Thus it vanishes in the limit $\epsilon \to 0$. Hence
\begin{equation}\label{formula}
f_{\rm exp}(y)=\lim_{\epsilon \to 0} \left[ -\frac{\epsilon}{2\pi y^2(1-y)}\sum_{n=1}^{\infty}I_n(y) \right] \;,
\end{equation}
where
\begin{equation}\label{In}
I_n(y)=\int_{n\pi-\pi/2}^{n\pi+\pi/2}dv \tanh^2\left(\frac{v}{2}\sqrt{\frac{y}{1-y}}\right)\frac{v\cos(v)-(3-2y)\sin(v)}{\sin^2(v)+\frac{\epsilon^2 v^2}{\left(2y\left(1-y\right)\right)^2} \cos^2(v)} \;.
\end{equation}
For $n \geq 1$, we need to keep the ${\cal O}(\epsilon^2)$ regulator in the denominator of the right hand side of Eq. (\ref{In}) since there is a double pole at $v=n \pi$. Therefore, in the $\epsilon \to 0$ limit, the dominant contribution to $I_n(y)$ comes from the neighbourhood of $v = n \pi$. Indeed, setting $v  =n \pi + \epsilon\, z$, we find  to leading order in the small $\epsilon$ limit 
\begin{equation}\label{In_2}
\begin{split}
I_n(y)\simeq\epsilon\int_{-\infty}^{+\infty}dz\tanh^2\left(\frac{n\pi+\epsilon z}{2}\sqrt{\frac{y}{1-y}}\right)\frac{(n\pi)(-1)^n}{\epsilon^2(z^2+\frac{(n\pi)^2}{(2y(1-y))^2})}=\frac{2\pi y(1-y)(-1)^n}{\epsilon}\tanh^2\left(\sqrt{\frac{y}{1-y}}\frac{n\pi}{2}\right).
\end{split}
\end{equation}
Substituting this result in Eq. (\ref{formula}), we see that the limit $\epsilon \to 0$ clearly exists and is given, for $0\leq y \leq 1$, by 
\begin{equation}\label{scaling_fexp}
f_{\rm exp}(y)=\frac{1}{y}\sum_{n=1}^{\infty}(-1)^{n-1}\tanh^2\left(\frac{n\pi}{2}\sqrt{\frac{y}{1-y}}\right) \;.
\end{equation}
This result above has been derived assuming $\tau = n_{\min} - n_{\max} > 0$, i.e., when the minimum occurs after the maximum. In the complementary case when $\tau < 0$ (when the maximum occurs after the minimum), it is clear that $P(\tau|n) = P(-\tau|n)$ and this follows simply from the time reversal symmetry of the process. Hence, we get, for $\tau \in [-n, n]$, and in the scaling limit $\tau, n \to \infty$ keeping the ratio $y = \tau/n$ fixed
\begin{eqnarray}\label{scaling_form2}
P(\tau | n) \underset{n \to \infty}{\longrightarrow} \frac{1}{n} f_{\rm exp} \left( \frac{\tau}{n}\right) \;.
\end{eqnarray}
where the scaling function $f_{\rm exp}(y)$ is given exactly by
\begin{eqnarray}\label{scaling_fexp2}
f_{\rm exp}(y)=\frac{1}{|y|}\sum_{n=1}^{\infty}(-1)^{n-1}\tanh^2\left(\frac{n\pi}{2}\sqrt{\frac{|y|}{1-|y|}}\right) \;.
\end{eqnarray}
Comparing with the Brownian case in Eq. (\ref{eq:f_bm}), we see that $f_{\rm exp}(y) = f_{\rm BM}(y)$. This exact computation for the double exponential jump distribution thus confirms explicitly the assertion of the CLT. 

\vspace*{0.3cm}

\noindent {\bf Direct proof of $f_{\rm exp}(y=1) = 1/2$}. We have seen above that the probability distribution $P(\tau|n)$ approaches a scaling form given in Eq. (\ref{scaling_form2}). The scaling function $f_{\rm exp}(y)$ is the same as the scaling function for the BM, $f_{\rm BM}(y)$ with asymptotic behaviors given in Eq. (\ref{summary_asymptotics}). In particular, we see that in the limit $y \to 1$, $f_{\rm BM}(y) \to 1/2$. As mentioned in the text, we have seen in our simulations that, for L\'evy flights with index $0<\mu \leq 2$, $P(\tau|n) \to (1/n) f_\mu(\tau/n)$ for large $n$ where the scaling function $f_\mu(y)$ depends on $\mu$. However, at the endpoints $f_\mu(y=\pm 1) = 1/2$ seems to be universal and independent of $\mu$ (see Fig. 3 of the main text). While it is difficult to prove analytically this universality at the endpoint for generic jump distributions including L\'evy flights, we show here that $f_{\rm exp}(y=1) = f_{\rm BM}(y=1)=1/2$ can be proved directly for the double exponential distribution $p(\eta) = (1/2) e^{-|\eta|}$. 

\begin{figure}[ht]
\includegraphics[width=0.6\linewidth]{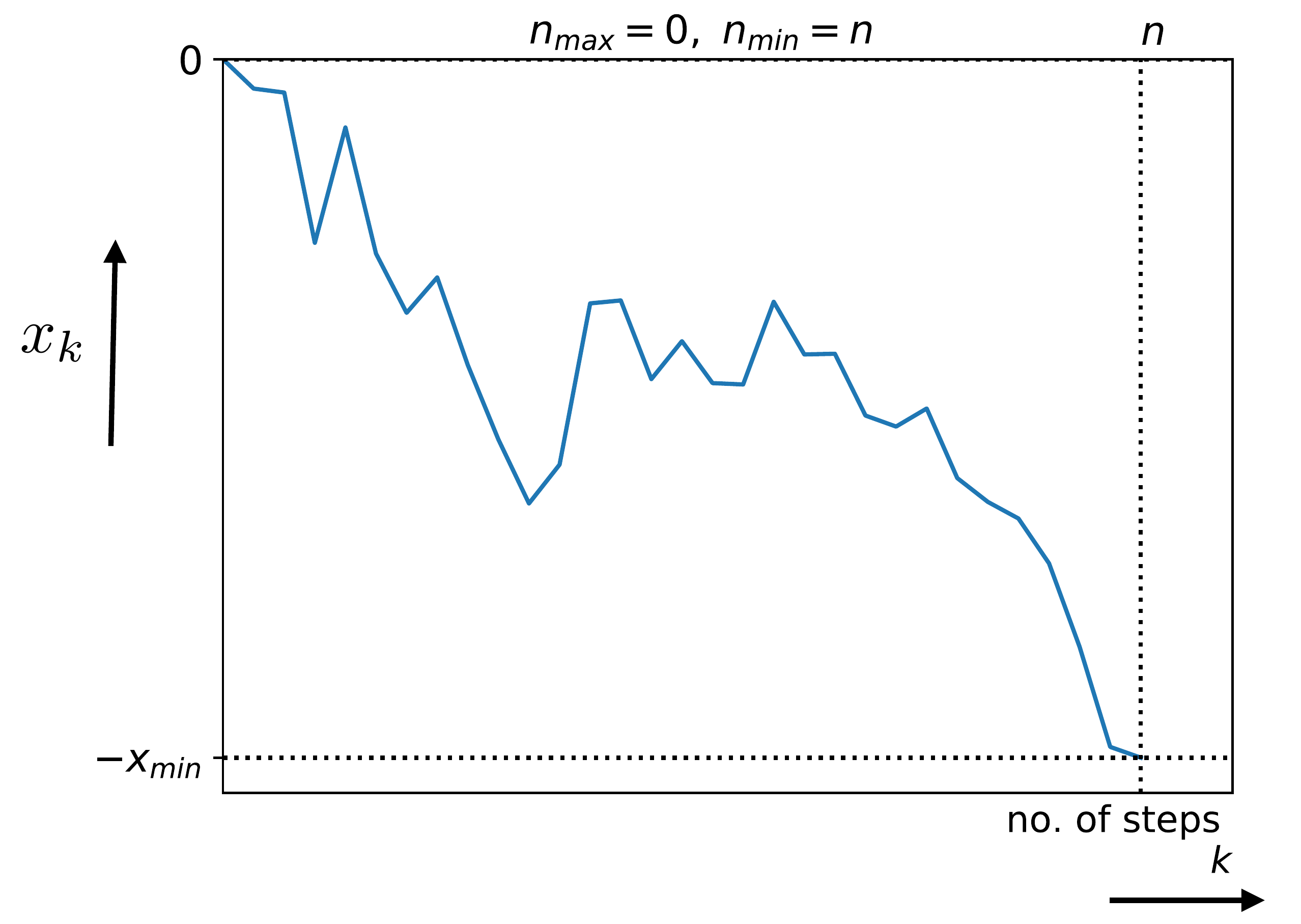}
\caption{A typical trajectory of a discrete-time random walk that contributes to the probability $P(\tau = n|n)$ where $\tau = n_{\min} - n_{\max}$. The event ''$\tau = n$'' can only happen when $n_{\max} = 0$ and $n_{\min} = n$, assuming $n_{\min} > n_{\max}$. Consequently, the trajectories that contribute to this event start at $x_0 = 0$ and arrive at $x_n = -x_{\min}$, while staying inside the box $[-x_{\min},0]$.}\label{Fig:1/2}
\end{figure}

The result $f_{\rm exp}(y=1) =1/2$ indicates that, for large $n$,  
\begin{eqnarray} \label{P1/2}
P(\tau = n_{\min} - n_{\max} = n|n) \approx \frac{1}{2\,n} \;. 
\end{eqnarray}
Since $-n \leq \tau \leq n$, it follows that the event ``$\tau = n_{\min} - n_{\max} = n$'' corresponds to having the maximum at step $k=0$ and the minimum at step $k=n$. This corresponds to a trajectory that starts at the origin at step $0$, reaches $-x_{\min}$ at step $n$, stays in the box $[-x_{\min},0]$ for all intermediate steps and finally one needs to integrate over all $x_{\min}$ in the range $[0, +\infty)$ (see Fig. \ref{Fig:1/2}). To compute the probability of such a trajectory, it is useful first to reflect the trajectory $x \to -x$. So now the probability of this event is just the probability that the trajectory starting at $0$ arrives at $x_{\min}\geq 0$ at step $n$, while staying in the box $[0,x_{\min}]$, with $x_{\min}$ integrated over $[0, +\infty)$. This probability can be conveniently expressed in terms of our basic building block $G(x,n|M)$ defined in Eq. (\ref{def_restricted}) as
\begin{eqnarray}\label{P1/2_2}
P(\tau  = n|n) = \int_0^\infty G(x_{\min}, n |x_{\min}) \, d x_{\min} \;.
\end{eqnarray}
Hence, to prove that $P(\tau = n |n) \approx 1/(2n)$, we need to evaluate the integral in the right hand side of Eq. (\ref{P1/2_2}) in the large $n$ limit. 
To evaluate this integral, we need the Green's function $G(x,n|M)$ for large $n$. Actually, for the double exponential jump distribution, the generating
function of $G(x,n|M)$ is given exactly in Eqs. (\ref{g_solution}) and (\ref{A}). This gives   
\begin{equation}\label{g_solution_bis}
 \tilde{G}\left(x_{\min},s|x_{\min} \right)=\sum_{n=1}^\infty {G}\left(x_{\min},n|x_{\min} \right)\, s^n =\frac{1-\sqrt{1-s}}{1-(\frac{1-\sqrt{1-s}}{1+\sqrt{1-s}})^2 e^{-2\sqrt{1-s}\,x_{\min}}}\left[e^{-\sqrt{1-s}\,x_{\min}}-\frac{1-\sqrt{1-s}}{1+\sqrt{1-s}}e^{-\sqrt{1-s}\,x_{\min}}\right].
\end{equation}
Since we are interested in the large $n$ behaviour, we need to investigate the $s\to 1$ limit of this expression. 
Setting $s=e^{-p}$ on the left hand side, the sum can be approximated by an integral in the limit $p\rightarrow 0$. Similarly, evaluating
the right hand side in the $p \to 0$ limit, we get, to leading order, 
\begin{equation}\label{P1/2_3}
\int_{0}^{\infty}dn \, G\left(x_{\min},n|x_{\min} \right)e^{-p n} = \frac{\sqrt{p}}{\sinh\left(x_{\min}\sqrt{p}\right)} \;.
\end{equation}
We next invert this Laplace transform with respect to $p$ using the identity 
\begin{equation}\label{eq_sinh}
\frac{\sqrt{p}}{\sinh\left(\sqrt{p}\right)}=\sum_{n=1}^{\infty}\frac{2n^2\pi^2(-1)^{n+1}}{p+n^2\pi^2} \;,
\end{equation}
and noting that each term on the right hand side corresponds to a simple pole in the complex $p$-plane. Hence the
inversion of the Laplace transform becomes simple and we get
\begin{equation}\label{eq_sinh_2}
G\left(x_{\min},n|x_{\min} \right)=2\pi^2\frac{1}{x_{\min}^3}\sum_{m=0}^{\infty}(-1)^{m+1}m^2e^{-\frac{m^2\pi^2}{x_{\min}^2}n}=\frac{1}{n}\frac{d}{dx_{\min}}\left[\sum_{m=0}^{\infty}(-1)^{m+1}e^{-\frac{m^2\pi^2}{x_{\min}^2}n}\right] \;.
\end{equation}
Integrating over $x_{\min}$, Eq. (\ref{P1/2_2}) gives
\begin{eqnarray}\label{P1/2_final}
P(\tau  = n|n) = \int_0^\infty G(x_{\min}, n |x_{\min}) d x_{\min} \approx  \frac{1}{n}\left(\sum_{m=0}^{\infty}(-1)^{m+1}+1\right)=\frac{1}{2n}\;.
\end{eqnarray}
Note that, in the last line, we have used the regularisation as in Eq. (\ref{yto1_2}) to evaluate the sum on the right hand side.

\section{Derivation of the distribution of $\tau = t_{\min} - t_{\max}$ for Brownian bridge}

\begin{figure}
\includegraphics[width=0.6\linewidth]{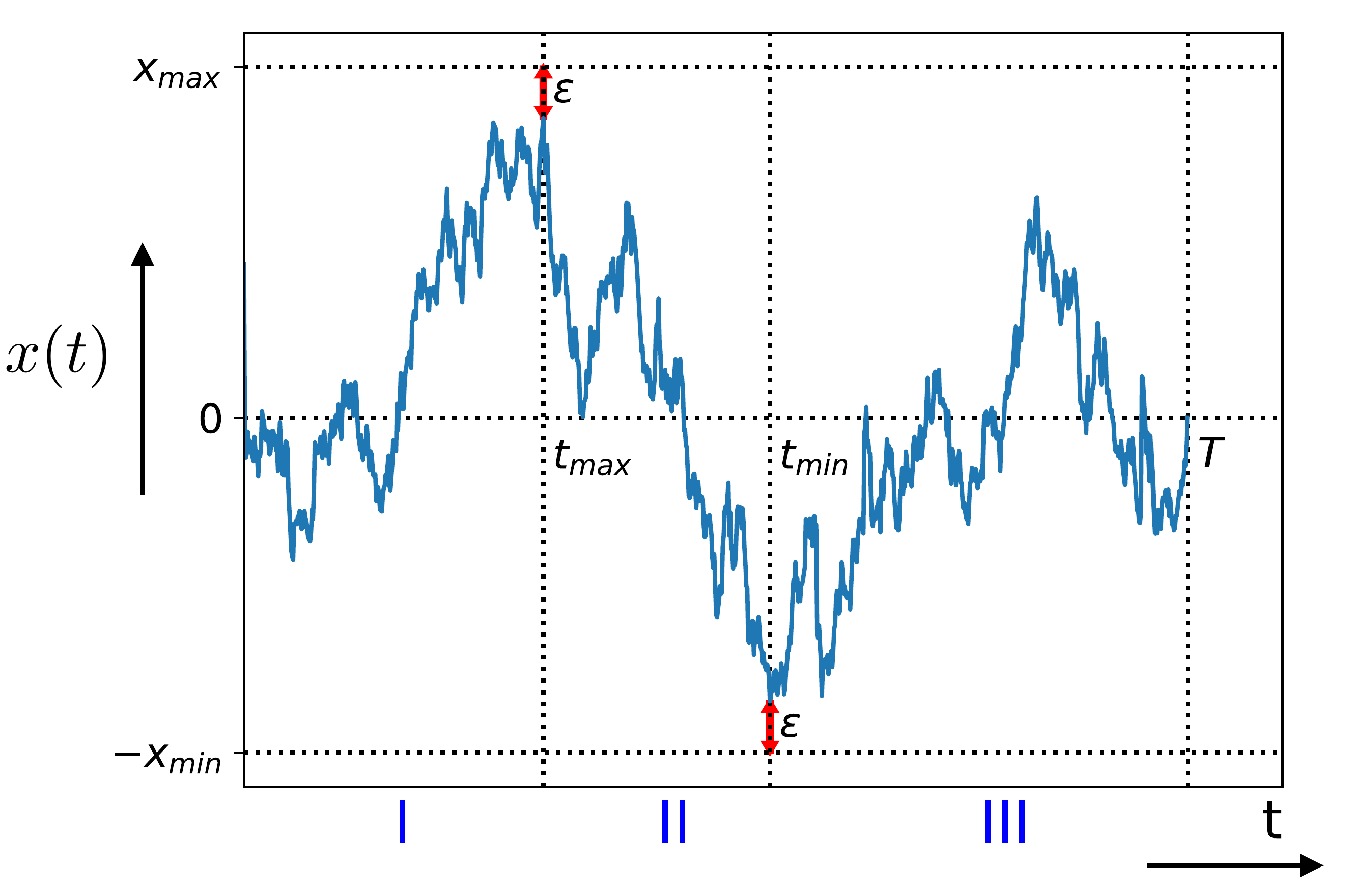}
\caption{A typical trajectory of a Brownian bridge $x(t)$ during the time interval $[0,T]$, starting from $x(0)=0$ and ending at $x(T)=0$. The value of the global maximum is $x_{\max} - \epsilon$ (with $\epsilon>0$) and the value of the global minimum $-x_{\min} + \epsilon$, where $\epsilon$ is the cut-off needed to enforce absorbing boundary conditions at $x_{\max}$ and $x_{\min}$ (as explained in the text). The time at which the maximum (respectively the minimum) occurs is denoted by $t_{\max}$ (respectively $t_{\min}$). The total time interval $[0,T]$ is divided into three segments: $[0,t_{\max}]$ (I), $[t_{\max}, t_{\min}]$ (II) and $[t_{\min}, T]$ (III), for the case $t_{\min}>t_{\max}$.}\label{Fig:BB}
\end{figure}

For the Brownian bridge (BB), one has the additional constraint that the Brownian motion, starting initially at $x_0=0$, comes back again
to the origin after time $T$ (see Fig. \ref{Fig:BB}). As in the case of the BM, we are interested in computing
the distribution $P(\tau|T)$ of the time difference $\tau = t_{\min} - t_{\max}$ between the occurrences of the minimum and the maximum. 
The result can be derived in two alternative ways: (i) by a direct path integral method as in the case of the BM and (ii) by using a mapping
between the BB and the Brownian excursion (BE) -- known as the Vervaat's construction in probability theory -- and then using the known results
for the BE. 

\vspace*{0.5cm}
\noindent{\bf Path-integral method.} The derivation follows more or less the same steps as in the BM case. We consider a typical trajectory
going from $x_0=0$ at time $t=0$ to the final position $x_{\rm f}=0$ at time $T$. Let $t_{\min}$ and $t_{\max}$ denote the time of occurrences 
of the minimum and the maximum respectively. The actual values of the minimum and the maximum are again denoted by $-x_{\min}$ and 
$x_{\max}$. As in the BM case, we first compute the grand joint PDF $P(x_{\min}, x_{\max}, t_{\min}, t_{\max} | T)$ by decomposing the interval $[0,T]$ into three segments I, II, III. While the probabilities $P_{\rm I}$ and $P_{\rm II}$ for the first two segments are exactly identical as in the BM case,
the probability for the last segment $P_{\rm III}$ is different, due to the bridge constraint $x_{\rm f}=0$. Once again, in terms of the Green's equation	
defined in Eq. (\ref{eq:g}), with the origin shifted to $-x_{\min}$ as in the BM case, this grand PDF can be written as
\begin{eqnarray} \label{jointPDF_BB}
P(x_{\min}, x_{\max}, t_{\min}, t_{\max} | T) \propto G_M(M-\epsilon, t_{\max}|x_{\min},0) \, G_M(\epsilon, t_{\min}|M-\epsilon, t_{\max}) \, G_M(x_{\min}-\epsilon, T|\epsilon, t_{\min}) \;,
\end{eqnarray}     
where we have again used the cut-off $\epsilon$, as explained in the BM case. Taking $\epsilon \to 0$ limit and integrating over $x_{\min}$ and $x_{\max}$, we can obtain the joint PDF $P(t_{\min}, t_{\max} | T)$. The intermediate steps leading to the final result are very similar to the BM case, hence we do not repeat them here and just quote the final result. For $t_{\min} > t_{\max}$, we get
\begin{eqnarray}\label{P>BB}
P_<(t_{\min}, t_{\max}|T) \propto B \sqrt{T} \sum_{n_1,n_2=1}^\infty \frac{(-1)^{n_1+n_2}\, n_1^2 n_2^2}{\left[n_1^2(T-(t_{\min}-t_{\max}))+ n_2^2 (t_{\min}-t_{\max})\right]^{5/2}} \, \theta(t_{\min}-t_{\max}) \;,
\end{eqnarray}
where the constant $B$ can be fixed from the overall normalisation. The factor $\sqrt{T}$ in Eq. (\ref{P>BB}) comes from the fact that, since we are considering a BB, we are implicitly conditioning on the event ``$x(T) = 0$''. Thus, after integrating out the variables $x_{\min}$ and $x_{\max}$ from the joint distribution in Eq. (\ref{jointPDF_BB}), one has also to divide by the probability 
\begin{equation}
P(x(T)=0|T)=\frac{1}{\sqrt{2\pi T}}\,,
\end{equation}
which gives the additional factor $\sqrt{T}$ in Eq. (\ref{P>BB}). We recall that the subscript '$<$' indicates $t_{\max} < t_{\min}$. To compute the PDF $P(\tau|T)$ of $\tau = t_{\min} - t_{\max}$, we focus on the case $t_{\min}>t_{\max}$, i.e. $\tau >0$. The complementary case $\tau<0$ is simply determined from the symmetry $P(-\tau|T) = P(\tau|T)$, as in the BM case. For $\tau >0$, one has
\begin{eqnarray}\label{Ptau_BB1}
P(\tau|T) = \int_0^T dt_{\max} \int_0^T dt_{\min} P_<(t_{\min}, t_{\max}|T) \; \delta(t_{\min} - t_{\max}-\tau) \;,
\end{eqnarray}
where $P_<(t_{\min}, t_{\max},|T)$ is given in Eq. (\ref{P>BB}). Noting that $P_<(t_{\min}, t_{\max},|T)$ depends only on the difference $\tau = t_{\min} - t_{\max}$, we can first carry out the integral over $t_{\min}$ in Eq. (\ref{Ptau_BB1}) keeping $\tau$ fixed. This gives an additional factor $(T-\tau)$ and we get, for $\tau > 0$
\begin{eqnarray}\label{Ptau_BB2}
P(\tau|T) = \frac{1}{T} f_{\rm BB}\left( \frac{\tau}{T}\right)
\end{eqnarray}
where the scaling function $f_{\rm BB}(y)$, for $0 \leq y \leq 1$, is given by
\begin{eqnarray}\label{fBB_SM1}
f_{\rm BB}(y) = B\;(1-y) \sum_{m,n=1}^{\infty}\frac{(-1)^{m+n}m^2n^2}{\left[m^2 \,y+n^2(1-y)\right]^{5/2}}  \;, \quad 0 \leq y \leq 1 \;.
\end{eqnarray}
For $\tau < 0$, using the symmetry $P(-\tau|T) = P(\tau|T)$, it follows that $P(\tau|T)$ takes exactly the same scaling form as in Eq. (\ref{Ptau_BB2}), with $y$ replaced by $-y$. Thus for all $-T \leq \tau \leq T$, $P(\tau|T) = (1/T) f_{\rm BB}(\tau/T)$ where $f_{\rm BB}(y)$, for all $-1 \leq y \leq 1$, is given by
\begin{eqnarray}\label{fBB_SM2}
f_{\rm BB}(y) = B\; (1-|y|) \sum_{m,n=1}^{\infty}\frac{(-1)^{m+n}m^2n^2}{\left[m^2 \,|y|+n^2(1-|y|)\right]^{5/2}} \;, \quad -1 \leq y \leq 1 \;.
\end{eqnarray}
Finally, the prefactor $B$ can be fixed from the normalisation condition $\int_{-1}^1 f_{\rm BB}(y) dy = 1$. By performing this integral, after a few steps of algebra, we find 
\begin{eqnarray} \label{A_BB}
B = 3 \;.
\end{eqnarray}
This then provides the derivation of the expression given in Eq. (6) of the main text. 

\begin{figure}[t]
 %\centering
\includegraphics[angle=0,width=\linewidth]{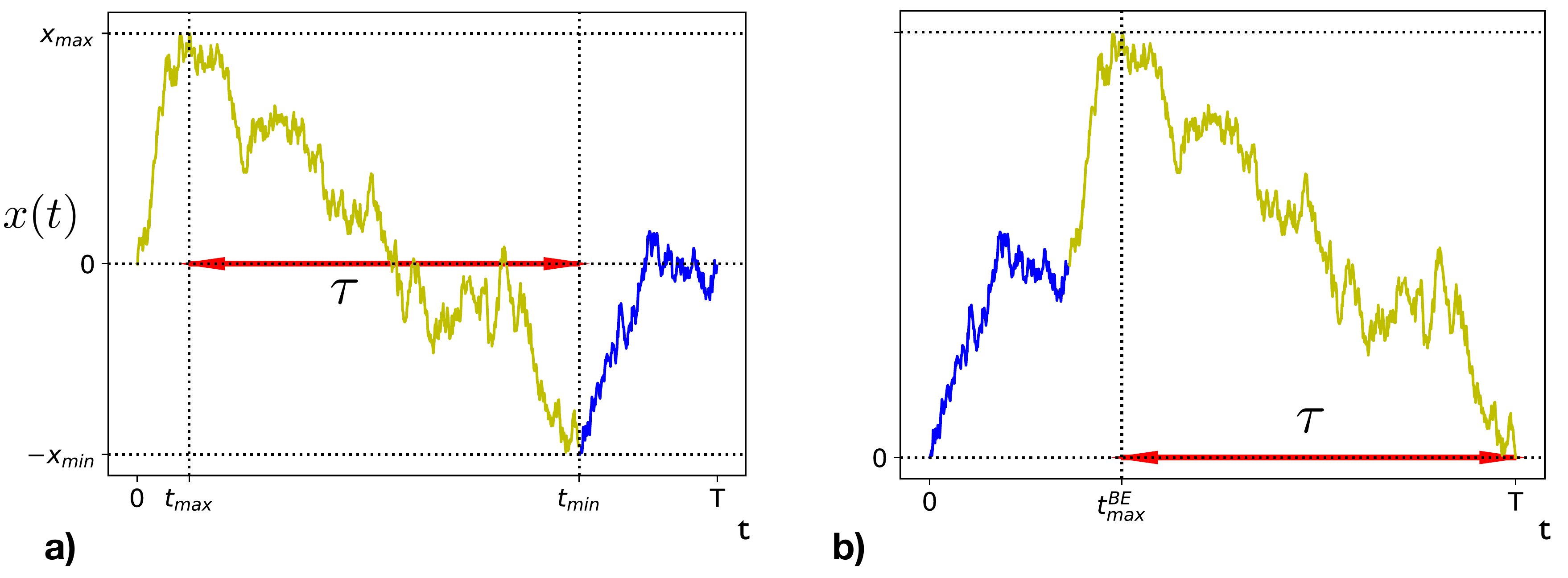}
\caption{Pictorial representation of the Vervaat transformation from a Brownian bridge $x(t)$ in panel a) to a Brownian excursion in panel b). On the left panel a), we have a Brownian bridge going from $x(0) = 0$ at time $t=0$ to the final position $x(T) = 0$ at time $t=T$. We first locate the time $t_{\min}$ at which the minimum of the bridge occurs with value $-x_{\min}$. We decompose the trajectory into two parts: the left of $t_{\min}$ (shown in yellow) and the right of $t_{\min}$ (shown in blue). Keeping the blue part of the trajectory fixed, we first slide forward the yellow part of the trajectory by an interval $T$ and glue this yellow part of the trajectory to the right end of the blue part. Next we shift the origin of the space to $-x_{\min}$. After these two transformations, the new trajectory on the right panel b) corresponds to a  Brownian excursion path. Note that the time difference $\tau = t_{\min} - t_{\max}$ in the bridge configuration on the left (indicated by a double arrowed red line) corresponds exactly to the time at which the maximum of the excursion occurs (measured from the right end the interval) on the right panel (also shown by a double arrowed red line).}
\label{fig:vervaat} 
\end{figure}

\vspace*{0.5cm}
\noindent{\bf Alternative derivation using Vervaat construction.} As mentioned before, we can derive the result for the BB in Eq. (\ref{fBB_SM2}) by using an alternative method based on a one-to-one mapping between a BB trajectory and a Brownian excursion (BE) trajectory -- known as Vervaat construction in probability theory \cite{vervaat79} (see also \cite{majumdar15}). Let us first recall that a BE on the time interval $[0,T]$ is a BB with the additional constraint that the path remains positive at all intermediate times between $0$ and $T$ (for a typical trajectory of BE, see the right panel of Fig. \ref{fig:vervaat}). From any BB configuration, one can obtain a BE configuration by sliding and fusing as explained in the caption of Fig. \ref{fig:vervaat}. Additionally, Vervaat proved that the configurations of BE generated by this construction from a BB configuration occur with the correct statistical weight corresponding to BE. Clearly, under this mapping, as also explained in the caption of Fig. \ref{fig:vervaat}, the time difference $\tau = t_{\min} - t_{\max}$ for a BB gets mapped onto $t^{\rm BE}_{\max}$ of a BE, measured from the right end of the interval $[0,T]$, where $t^{\rm BE}_{\max}$ denotes the time at which the maximum of a BE occurs. This is a random variable, and let us denote its PDF by
\begin{eqnarray} \label{def_PBE}
P_{\rm BE}(\tau|T) = {\rm Prob.}\left(t_{\max}^{\rm BE} = \tau|T\right) \;.
\end{eqnarray}
Note that this mapping is one-to-one only if we fix the value of $t_{\min}$. Thus, focusing on the case where $\tau >0$, i.e. $t_{\min} > t_{\max}$ for BB, the Vervaat construction provides the exact identity
\begin{equation}\label{eq:BB_BE}
P_{\rm BB}(t_{\min}-t_{\max}|t_{\min},T)  = P_{\rm BE}(t_{\min}-t_{\max}|T)   \;,
\end{equation}
where the left-hand side denotes the PDF of $t_{\min}-t_{\max}$ for a BB, conditioned on $t_{\min}$ and on the total time $T$. For the BE, the PDF $P_{\rm BE}(\tau|T)$ was computed exactly in \cite{randon-furling08} and it reads
\begin{eqnarray}\label{PBE_explicit}
P_{\rm BE}(\tau|T)=3\, T^{3/2} \,\sum_{m,n=1}^{\infty} \frac{(-1)^{m+n}m^2n^2 }{\left[m^2 \tau+n^2(T-\tau)\right]^{5/2}} \;.
\end{eqnarray}
The joint PDF of $t_{\max}$ and $t_{\min}$ for a BB can be written as
\begin{eqnarray}\label{eq:joint_BB}
P_{\rm BB}(t_{\max},t_{\min}|T)&=& P_{\rm BB}(t_{\max}-t_{\min}|t_{\min},T)P_{\rm BB}(t_{\min}|T)
=P_{\rm BE}(t_{\min}-t_{\max}|T) P_{\rm BB}(t_{\min}|T)\,,
\end{eqnarray}
where we have used Eq. (\ref{eq:BB_BE}). The distribution of the time $t_{\min}$ of the minimum of a BB is uniform over $[0,T]$ \cite{morters10}
\begin{equation}\label{eq:prob_tmin}
P_{\rm BB}(t_{\min}|T)=\frac1T\,,
\end{equation}
Thus, plugging the expressions for $P_{\rm BE}(t_{\min}-t_{\max}|T)$ and $P_{\rm BB}(t_{\min}|T)$, given in Eqs. (\ref{PBE_explicit}) and (\ref{eq:prob_tmin}), into Eq. (\ref{eq:joint_BB}), we obtain
\begin{eqnarray}\label{eq:P_tmax_tmin_BB}
&& P_{\rm BB}(t_{\max},t_{\min}|T)=
3\, \sqrt{T} \sum_{m,n=1}^{\infty} \frac{(-1)^{m+n}m^2n^2 }{\left[m^2 (t_{\min}-t_{\max})+n^2(T-t_{\min}+t_{\max})\right]^{5/2}}\,,
\end{eqnarray}
Finally, integrating Eq. (\ref{eq:P_tmax_tmin_BB}) over $t_{\max}$ and $t_{\min}$, keeping $\tau=t_{\min}-t_{\max}$ fixed, we obtain exactly the same scaling form as in Eq. (\ref{Ptau_BB2}) with the same scaling function $f_{\rm BB}(y)$ obtained in Eq. (\ref{fBB_SM2}) by the path-integral method.

\section{Covariance of $t_{\min}$ and $t_{\max}$}

We want to compute the covariance between the time of the minimum $t_{\min}$ and the time of the maximum $t_{\max}$ of the BM as well as the BB. By definition, 
\begin{equation} \label{def_cov}
\operatorname{cov}\left(t_{\min},t_{\max}\right)= \langle t_{\min} t_{\max}\rangle  -\langle t_{\min}\rangle  \langle t_{\max}\rangle  \;.
\end{equation}
It is first convenient to express this covariance in terms of the variable $\tau = t_{\min} - t_{\max}$ by noting that 
\begin{eqnarray}\label{id_tau}
\langle \tau^2 \rangle = \langle t_{\min}^2 \rangle + \langle t_{\max}^2 \rangle - 2 \langle t_{\min} \, t_{\max} \rangle  \;.
\end{eqnarray}
Hence, 
\begin{equation}\label{formula_covariance}
\operatorname{cov}\left(t_{\min},t_{\max}\right)=\frac{1}{2}\left(\langle t_{\min}^2\rangle +\langle t_{\max}^2\rangle -\langle \tau^2\rangle \right)-\langle t_{\min}\rangle \langle t_{\max}\rangle  \;.
\end{equation}
Thus we need the statistics of $t_{\min}$, $t_{\max}$ and $\tau$. In the case of the BM, it was already mentioned in the main text that the marginal PDFs of $t_{\min}$ and $t_{\max}$ are given by the derivative of the arcsine law of their respective cumulative distributions,   
\begin{eqnarray} 
&&P(t_{\min}|T)=\frac{1}{\pi\sqrt{t_{\min}(T-t_{\min})}} \;, \;\; \quad \; 0\leq t_{\min} \leq T \;, \label{arcsine_min} \\
&&P(t_{\max}|T)=\frac{1}{\pi\sqrt{t_{\max}(T-t_{\max})}} \;, \; \quad 0\leq t_{\max} \leq T \;. \label{arcsine_max} \;.
\end{eqnarray}
From Eqs. (\ref{arcsine_min}) and  (\ref{arcsine_max}) we get
\begin{eqnarray}
&&\langle t_{\min} \rangle = \langle t_{\max} \rangle = \frac{T}{2} \label{av_tmin} \;, \\
&&\langle t_{\min}^2 \rangle = \langle t_{\max}^2 \rangle = \frac{3}{8}T^2 \;. \label{var_tmin}
\end{eqnarray}
It rests to compute $\langle \tau^2 \rangle$. Using the fact that $P(\tau | T) = (1/T) f_{\rm BM}(\tau/T)$ for $-T \leq \tau \leq T $, we get 
\begin{equation} \label{var_tau1}
\langle \tau^2\rangle = \int_{-T}^T d\tau \, \tau^2 \, \frac{1}{T} f_{\rm BM}\left( \frac{\tau}{T}\right) = 2T^2\int_{0}^{1}dy \, y^2 \, f_{\rm BM}(y) \;.
\end{equation}
Computing this integral using directly the expression of $f_{\rm BM}(y)$ given in Eq. (\ref{eq:f_bm}) is hard. It is actually more convenient to use the integral equation satisfied by $f_{\rm BM}(y) = f_{\rm exp}(y)$ given in Eq. (\ref{integral2}). The last integral can be computed from Eq. (\ref{integral2}), by differentiating twice with respect to $u$ and by setting $u=0$
\begin{eqnarray} 
\int_{0}^{1}dy \, y^2 \, f_{\rm BM}(y)&=&\int_{0}^{\infty}dz\frac{e^{-z}}{1-e^{-2z}}\frac{\partial^2}{\partial u^2}\left[\tanh^2\left(\frac{z}{2\sqrt{1+u}}\right)\right]\biggr\rvert_{u=0} \nonumber \\
&=& \frac{1}{32} \int_0^\infty dz\, z \, \text{csch}\left(\frac{z}{2}\right) \text{sech}^5\left(\frac{z}{2}\right) \left[2 z+3 \sinh (z) -z \,\cosh (z)\right]  \nonumber \\
&=&\frac{7\zeta(3)-2}{32} \;. \label{var_tau2}
\end{eqnarray}
Substituting the results from Eqs. (\ref{var_tau2}) (\ref{av_tmin}) and (\ref{var_tmin}) in Eq. (\ref{formula_covariance}), we get 
\begin{equation} \label{cov}
\operatorname{cov}_{BM}(t_{\min},t_{\max})=-\frac{7\zeta(3)-6}{32}T^2\simeq -(0.0754\ldots) \, T^2.
\end{equation}

For the BB, the same formula (\ref{formula_covariance}) holds. The marginal PDFs of $t_{\min}$ and $t_{\max}$ are both uniform over $[0,T]$ \cite{morters10}
\begin{eqnarray} 
&&P(t_{\min}|T)=\frac{1}{T} \;, \;\; \quad \; 0\leq t_{\min} \leq T \;, \label{unif_min} \\
&&P(t_{\max}|T)=\frac{1}{T} \;, \; \quad 0\leq t_{\max} \leq T \;. \label{unif_max} 
\end{eqnarray}
This gives the first two moments
\begin{eqnarray}
&&\langle t_{\min} \rangle = \langle t_{\max} \rangle = \frac{T}{2} \label{av_tmin_BB} \;, \\
&&\langle t_{\min}^2 \rangle = \langle t_{\max}^2 \rangle = \frac{1}{3}T^2 \;. \label{var_tmin_BB} 
\end{eqnarray}
To compute $\langle \tau^2 \rangle$ we use the scaling form $P(\tau|T) = (1/T) f_{\rm BB}(\tau/T)$. Therefore 
\begin{eqnarray} \label{var_tau_BB}
\langle \tau^2 \rangle = \int_{-T}^T d\tau \, \tau^2 \, \frac{1}{T} f_{\rm BB} \left( \frac{\tau}{T}\right) = 2 T^2 \int_0^1 dy \, y^2 \, f_{\rm BB}(y) \;,
\end{eqnarray}
where the scaling function $f_{\rm BB}(y)$ is given in Eq. (6) in the main text. As argued before, the scaling function can be related to the PDF of $t_{\max}$ of a Brownian excursion over the interval $[0,T]$. Indeed, $P_{\rm BE}(t_{\max}|T) = (1/T) f_{\rm BE}(t_{\max}/T)$ and the two scaling functions $f_{\rm BB}(y)$ and $f_{\rm BE}(y)$, for $0\leq y \leq 1$, are simply related by
\begin{eqnarray} \label{rel_BB_BE}
f_{\rm BB}(y) = (1-y) f_{\rm BE}(y) \;,
\end{eqnarray}
where $f_{\rm BE}(y)$ is given by \cite{randon-furling08}
\begin{eqnarray} \label{fBE}
f_{\rm BE}(y) = 3 \sum_{m,n=1}^{\infty}\frac{(-1)^{m+n}m^2n^2}{\left(m^2 y+n^2(1-y)\right)^{5/2}} \;.
\end{eqnarray}
Consequently, from Eq. (\ref{var_tau_BB}), we get
\begin{eqnarray}\label{var_tau_BB2}
\langle \tau^2 \rangle = 2  T^2 \, \Big( \langle y^2\rangle_{\rm BE} - \langle y^3\rangle_{\rm BE} \Big)\,,
\end{eqnarray}
where
\begin{eqnarray}\label{def_mom_BE}
\langle y^m \rangle_{\rm BE} = \int_0^1 dy \, y^m \, f_{\rm BE}(y) \;. 
\end{eqnarray}
The first of these three moments for $m=1,2,3$ were computed in Ref. \cite{randon-furling08}
\begin{equation}
\left\langle y^2\right\rangle _{\rm BE}= \frac{1}{2} ,\textit{        }\textit{        }\textit{        } \left\langle y^2\right\rangle _{\rm BE}=\frac{15-\pi^2}{18},\textit{        }\textit{        }\textit{        }\left\langle y^3\right\rangle _{\rm BE}=1-\frac{\pi^2}{12}.
\end{equation}
Substituting these results in Eq. (\ref{var_tau_BB2}) and further using the results from Eqs. (\ref{av_tmin_BB}) and (\ref{var_tmin_BB}) in Eq. (\ref{formula_covariance}) finally gives
\begin{equation} \label{final_cov_BB}
\operatorname{cov}_{BB}(t_{\min},t_{\max})=-\frac{\pi^2-9}{36}T^2 = - (0.0241 \ldots) T^2\;.
\end{equation}
Thus, by comparing Eqs. (\ref{cov}) and (\ref{final_cov_BB}), we see that $t_{\min}$ and $t_{\max}$ are more strongly anti-correlated in the BM case than the BB case.

\section{Application to fluctuating interfaces in $1+1$ dimensions}

In this section, we demonstrate how our results can be applied to $(1+1)$-dimensional fluctuating interfaces of
the Kardar-Parisi-Zhang (KPZ) or Edwards-Wilkinson (EW) variety. We consider a one-dimensional interface
growing on a finite substrate of length $L$ and denote by $H(x,t)$ the height of the interface at position $x$ at time
$t$. The height field evolves by the KPZ equation 
\begin{equation} \label{eq:kpz_sup}
\frac{\partial H(x,t)}{\partial t}=\frac{\partial^2 H(x,t)}{\partial x^2}+\lambda\left(\frac{\partial H(x,t)}{\partial x}\right)^2+\eta\left(x,t\right),
\end{equation}
where $\lambda \geq 0$ and $\eta(x,t)$ is a Gaussian white noise with zero mean and correlator $\left<\eta\left(x,t\right)\eta\left(x',t'\right)\right>=2\delta(x-x')\delta(t-t')$. When $\lambda = 0$, this equation (\ref{eq:kpz_sup}) becomes linear and is known as the EW equation. We consider both (i) free boundary conditions (FBC) where $H(x=0,t)$ and $H(x=L,t)$ evolve freely and (ii) periodic boundary conditions (PBC) where $H(x=0,t)$ and $H(x=L,t)$ evolve freely as in (i), but with an additional constraint $H(x=0,t) = H(x=L,t)$. As argued in the main text, we subtract the zero mode  $\overline{H(t)}=(1/L)\int_{0}^{L}H(x,t)\,dx$, and focus on the relative height $h(x,t) = H(x,t) - \overline{H(t)}$, whose distribution reaches a stationary state in the long time limit, in a finite system. Let $h_{\min}$ and $h_{\max}$ denote the maximal and the minimal relative height. In a given realisation, let $h_{\min}$ (respectively $h_{\max}$) occur at position $\tilde x_{\min}$ (respectively $\tilde x_{\max}$). Our main objects of interest are the joint PDF of $\tilde x_{\min}$ and $\tilde x_{\max}$ denoted by $P(\tilde x_{\min}, \tilde x_{\max}|L)$, and in particular the PDF of the position difference between the minimum and the maximum $\tau = \tilde x_{\min} - \tilde x_{\max}$, which we will denote by $P(\tau|L)$.    

\vspace*{0.3cm}

\noindent {\bf Edwards-Wilkinson case}. We start with the simpler case of the EW interface where $\lambda = 0$ in Eq. (\ref{eq:kpz_sup}). In this case, as argued in the text, for the FBC, 
the relative heights at different spatial points reach a stationary state {\it for any finite $L$}, with the joint PDF given by 
\begin{equation}\label{eq:stationary_sup}
P_{\rm st}\left(\{h\}\right)=B_L\,e^{-1/2\int_{0}^{L}dx (\partial_{x}h)^2}\delta\left[\int_{0}^{L}h(x)dx\right] \;,
\end{equation}
where $B_L$ is a normalization constant. The delta-function imposes the constraint that area under the interface 
is identically zero since $h(x,t)$ corresponds to the {\it relative} height. In contrast, for the PBC, the stationary joint PDF of the relative heights, for any finite $L$, reads
\begin{equation}\label{eq:stationary_pbc_sup}
P_{\rm st}\left(\{h\}\right)=A_L\,e^{-1/2\int_{0}^{L}dx (\partial_{x}h)^2}\delta\left[\int_{0}^{L}h(x)dx\right] \, \delta[h(L)-h(0)] \;,
\end{equation}
with $A_L$ the corresponding normalisation constant. In both cases, it is clear from the stationary measures in Eqs. (\ref{eq:stationary_sup}) and (\ref{eq:stationary_pbc_sup}) that, apart from the global zero-area constraint and the boundary conditions, the stationary interface behaves as a Brownian motion locally in space. Thus, identifying (a) space with time, i.e., $x \Leftrightarrow t$, (b) the total substrate length $L$ with the total duration $T$, i.e., $L \Leftrightarrow T$ and (c) the stationary interface height $h(x)$ with the position $x(t)$ of a Brownian motion (BM), i.e., $h(x) \Leftrightarrow x(t)$, we find a one-to-one mapping between the stationary EW interface and the positions of a BM (for the case of the FBC). In the case of the PBC, the stationary interface corresponds to a Brownian bridge (BB). There is however an important difference: the BM and BB thus obtained from the stationary interface satisfy a global constraint $\int_0^T x(t)\, dt = 0$, i.e. the area under the curve is identically zero. While this constraint affects the actual values of the heights and their distributions, the positions at which the minimum or the maximum of the relative height occurs do not depend on this zero mode, both for the FBC and the PBC. Hence, for the FBC, 
using the fact that the process is locally Brownian, we expect that the joint distribution $P(\tilde x_{\min}, \tilde x_{\max}|L)$ will coincide with the corresponding joint PDF of $t_{\min}$ and $t_{\max}$ of the BM with duration $T=L$. Consequently, the distribution of the position difference between the minimum and the maximum relative height in the EW stationary interface with the FBC will be given by
\begin{eqnarray}\label{Ptau_FBC}
P(\tau = \tilde x_{\min} - \tilde x_{\max}|L) = \frac{1}{L} f_{\rm BM}\left( \frac{\tau}{L}\right) \;,
\end{eqnarray}   
where the scaling function $f_{\rm BM}(y)$ is given in Eq. (4) of the main text [see also Eq. (\ref{Poisson4}) in this Supplementary Material]. For the case of the PBC, again using the local Brownian nature of the stationary EW interface, it follows that the PDF of the position difference will converge to the case of a BB
\begin{eqnarray}\label{Ptau_PBC}
P(\tau = \tilde x_{\min} - \tilde x_{\max}|L) = \frac{1}{L} f_{\rm BB}\left( \frac{\tau}{L}\right) \;,
\end{eqnarray}     
where $f_{\rm BB}(y)$ is given exactly in Eq. (6) of the main text [see also Eq. (\ref{fBB_SM2}) in this Supplementary Material]. We note that the results in Eqs. (\ref{Ptau_FBC}) and (\ref{Ptau_PBC}) are valid for any finite $L$ in the stationary state (and not just asymptotically for large $L$). 

\begin{figure}[ht]    
 \includegraphics[angle=0,width=\linewidth]{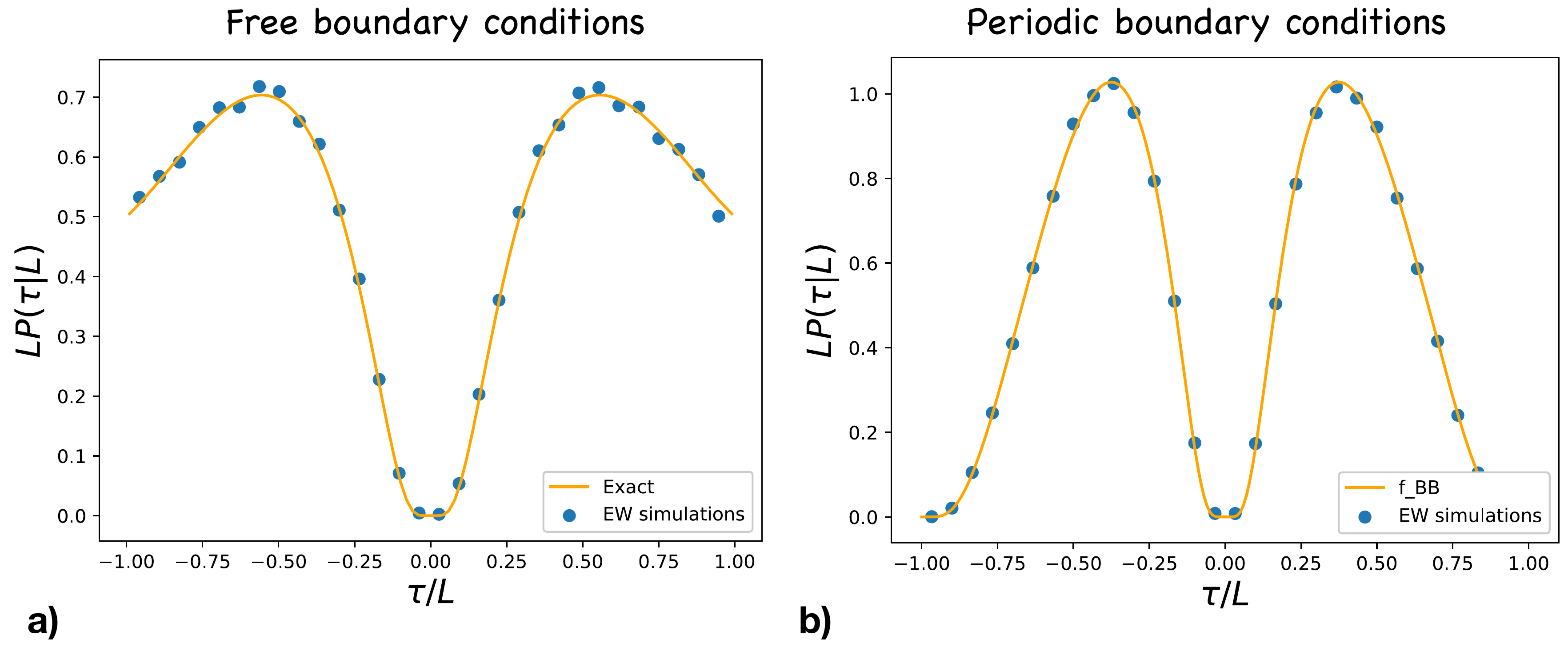} 
    \caption{
Scaling plot of $P(\tau = \tilde x_{\min} - \tilde x_{\max}|L)$ for the EW interface obtained from the numerical integration of Eq. (\ref{discrete_EW}) with $\Delta t=0.01$ and $L=512$: a) for the FBC and b) for the PBC. The solid line in a) represents the analytical scaling function $f_{\rm BM}(y)$ given in Eq. (\ref{Poisson4}) while the filled dots represent simulation data. In b), the solid line represents the analytical scaling function $f_{\rm BB}(y)$ given in Eq. (\ref{fBB_SM2}), while the filled dots represent simulation data. The numerical data are obtained by averaging over $10^6$ samples.}
    \label{fig:numeric_ew}
\end{figure}
To check this prediction for $P(\tau|L)$ in Eqs. (\ref{Ptau_FBC}) and (\ref{Ptau_PBC}) for the EW interface, we numerically integrated the space-time discretised form of Eq. (\ref{eq:kpz_sup}) with $\lambda = 0$
\begin{eqnarray}\label{discrete_EW}
H(i,t+ \Delta t) - H(i,t) = \Delta t \left[H(i+1,t) + H(i-1,t) - 2 H(i,t) \right] + \eta_i(t) \sqrt{2 \Delta t} \;,
\end{eqnarray}
where $\eta_i(t)$'s are independent and identically distributed random variables for each $i$ and $t$ and each drawn from a Gaussian distribution of zero mean and unit variance. We considered both the FBC and the PBC with $\Delta t = 0.01$ and $L = 512$. We have run the simulation for a sufficiently large time to ensure that the system has reached the stationary state and then measured the PDF $P(\tau|L)$. Even though we expect the results in  Eqs. (\ref{Ptau_FBC}) and (\ref{Ptau_PBC}) to be valid for all $L$, this expectation is only for the continuum version of the EW equation (\ref{eq:kpz_sup}) with $\lambda=0$. Since for the simulation we have discretised this equation, we expect these results in Eqs. (\ref{Ptau_FBC}) and (\ref{Ptau_PBC}) to hold only for large $L$. Actually, for $L=512$, we already see an excellent agreement between simulations and analytical results. In Fig. \ref{fig:numeric_ew} a) we compare the simulations with the analytical prediction for the FBC in Eq. (\ref{Ptau_FBC}). The corresponding simulation results for the PBC are shown in Fig. \ref{fig:numeric_ew} b) and compared with the analytical prediction in Eq. (\ref{Ptau_PBC}).

\vspace*{0.3cm}

\noindent {\bf The KPZ case}. As mentioned in the main text, for the KPZ equation (\ref{eq:kpz_sup}) with $\lambda >0$, the stationary state for the relative heights is expected to converge to the same measures (\ref{eq:stationary_sup}) and (\ref{eq:stationary_pbc_sup}) (respectively for the FBC and the PBC), but only in the limit $L \to \infty$. Therefore, we would expect that the results for $P(\tau|L)$ in Eqs. (\ref{Ptau_FBC}) and (\ref{Ptau_PBC}) to hold also for the KPZ equation for large $L$. However, verifying these analytical predictions numerically for the KPZ equation is challenging because the non-linear term is not easy to discretize \cite{lam_shin,lam_shin2}. Several discretisation schemes have been proposed in the literature and we found it suitable to use the scheme proposed by Lam and Shin \cite{lam_shin2}, where the non-linear term $\lambda (\partial_x H(x,t))^2$ is discretised as follows
\begin{eqnarray}\label{lam_shin}
\frac{\lambda}{3} \left[ (H(i+1,t)-H(i,t))^2 + (H(i+1,t)- H(i,t)) (H(i,t) - H(i-1,t)) + (H(i,t) - H(i-1,t))^2  )\right] \;.
\end{eqnarray} 
\begin{figure}[ht]    
 \includegraphics[width=0.55\linewidth]{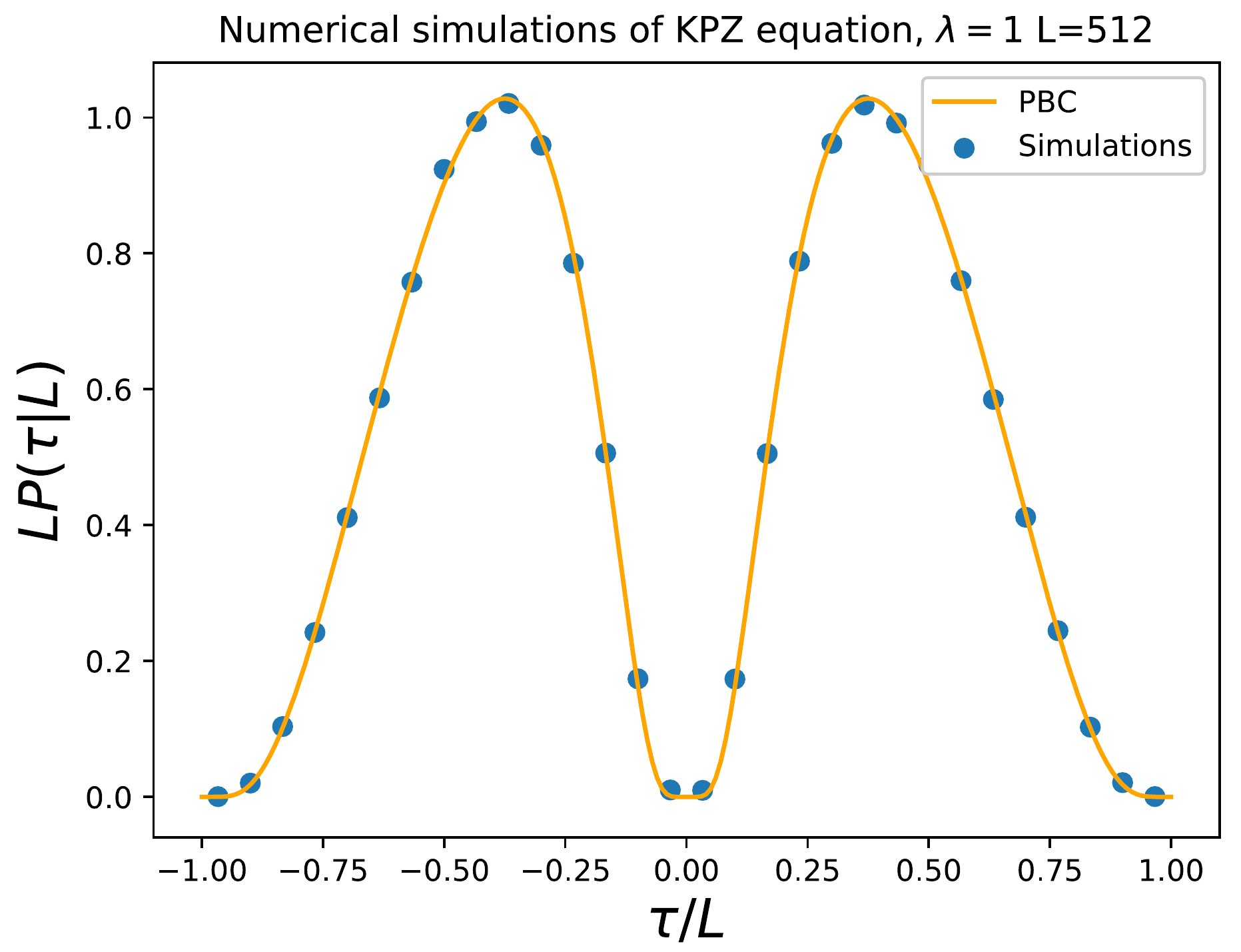} 
    \caption{Scaling plot of $P(\tau = \tilde x_{\min} - \tilde x_{\max}|L)$ for the KPZ interface with PBC obtained using the discretisation scheme (\ref{lam_shin}) with
    $\Delta t=0.01$ and $L=512$. The solid line represents the analytical scaling function $f_{\rm BB}(y)$ given in Eq. (\ref{fBB_SM2}), while the filled dots represent the simulation data. The numerical data are obtained by averaging over $10^6$ samples.}\label{Fig_KPZ}
\end{figure}    
The advantage of this scheme is that one can prove analytically that, for the PBC, the Fokker-Planck equation associated with this discrete model admits a stationary solution, $P_{\rm st}(\{ H\}) \propto \exp \left[-\frac{1}{2} \sum_{i=1}^L (H(i+1,t)-H(i,t))^2 \right]$, independently of $\lambda$. In the $L \to \infty$ limit, the stationary measure converges to the Brownian measure  $P_{\rm st}(\{ H\}) \propto \exp \left[-\frac{1}{2} \int_0^L (\partial_x H)^2 \, dx \right]$. Therefore, with this discretisation scheme (\ref{lam_shin}) and PBC, we expect to recover the Brownian bridge result for $P(\tau|L)$ as in Eq. (\ref{Ptau_PBC}). In Fig. \ref{Fig_KPZ} we compare the simulation results for $P(\tau|L)$ for the KPZ equation with PBC and $\lambda = 1$ (with parameter $\Delta t = 0.01$ and $L=512$), with the analytical scaling function in Eq. (\ref{Ptau_PBC}) for the Brownian bridge -- the agreement is excellent. Unfortunately, for the KPZ equation with the FBC, there is no convenient discretisation scheme for the non-linear term that correctly produces the stationary measure for finite $L$. Of course, we still expect that, in this case, the results for $P(\tau|L)$ for the KPZ equation in the stationary state will again converge to the Brownian prediction given in Eq. (\ref{Ptau_FBC}) in the large $L$ limit. However, numerically verifying this for finite but large $L$ seems challenging, due to the absence of a good discretisation scheme for the non-linear term in the FBC case.